\documentstyle[12pt,epsfig,fleqn]{article}
\newcommand{\mybox}{\mbox}
\newcommand{\be}{\begin{eqnarray}}
\newcommand{\ee}{\end{eqnarray}}
\newcommand{\nee}{\nonumber\end{eqnarray}}

\newcommand{\msb}[1]   {m_{\sb_{#1}}}

\newcommand{\msg}      {m_{\ti g}}

\def\lag             {{\cal L}}

\def\R               {{\cal R}}

\def\PL              {P_L^{}}
\def\PR              {P_R^{}}

\def\rzw             {\sqrt{2}}
\def\delr            {\!\stackrel{\leftrightarrow}{\partial^\mu}\!}
\newcommand{\gsim}{\;\raisebox{-0.9ex}
           {$\textstyle\stackrel{\textstyle >}{\sim}$}\;}
\newcommand{\lsim}{\;\raisebox{-0.9ex}{$\textstyle\stackrel{\textstyle<}
           {\sim}$}\;}

\def\a               {\alpha}
\def\b               {\beta}

\def\g               {\gamma}

\def\ti              {\tilde}
\def\sq              {\ti q}
\def\st              {\ti t}
\def\sb              {\ti b}

\def\sg              {\ti g}

\title{CP violation in  $H^\pm t$ production at the LHC}

\author{E. Christova$^{1}$, H. Eberl$^{2}$,
        E. Ginina$^{2}$, W. Majerotto$^{2}$}
\date{
\small $^{1}$Institute for Nuclear Research and Nuclear Energy, BAS, Sofia 1784, Bulgaria\\
    $^{2}$Institut f\"ur Hochenergiephysik der \"Osterreichischen Akademie der
     Wissenschaften,  A-1050 Vienna, Austria\\
     E-mails: echristo@inrne.bas.bg, helmut@hephy.oeaw.ac.at,\\
     eginina@hephy.oeaw.ac.at, majer@hephy.oeaw.ac.at}

\makeindex

\begin{document}

\maketitle

\begin{abstract}
We study effects of CP violation in the associated production of a
charged Higgs boson and a top quark at the LHC, $pp \to tH^\pm
+X$. We calculate the CP violating asymmetry between the total
cross section for $H^+$ and $H^-$ production at next-to-leading
order in the minimal supersymmetric standard model (MSSM), and perform a detailed numerical analysis.  In
the production the asymmetry is of the order of 20\%. The
asymmetry in the production and any subsequent decay of an
on-shell charged Higgs boson is to a good approximation the sum of
the asymmetry in the production and the asymmetry in the decay. We
consider subsequent decays of $H^\pm $ to $tb$, $\nu_\tau
\tau^\pm$ and $Wh^0$. In the case with $H^\pm \to tb$ decay, mainly due to CP violating box graphs with
gluino, the asymmetry can go up to $\sim$~12\%.
\end{abstract}

\clearpage

\tableofcontents

\section{Introduction}

If a charged $H^\pm$ boson is discovered at LHC  or at any future
collider, it would be a clear signal for Physics beyond the
Standard Model (SM). The next question would be which Physics
beyond the SM it is -- almost all extensions of the SM contain a
larger Higgs sector and inevitably predict the
existence of a charged Higgs boson. CP violation (CPV) is
a possible tool to disentangle the different charged Higgs
bosons. The phenomena of CPV is important also because it is
believed that this is the key to our understanding the observed
abundance  of matter over antimatter. Most extensions of the SM
contain possible new sources of CPV through additional CPV phases.

In this note we study CPV in the Minimal Supersymmetric
Standard Model (MSSM) with complex couplings, being one of the
most promising candidates for an extension of the SM. In
 MSSM the additional sources of CPV are the phases of
 the higgsino mass parameter $\mu = \vert \mu\vert e^{i\phi_\mu}$ in the superpotential,
 of the
 gaugino mass parameters $M_i=\vert M_i\vert e^{i\phi_i}$, $i=1,2,3$ and of the
trilinear couplings $A_f=\vert A_f\vert e^{i\phi_f}$
(corresponding to a fermion $f$)~\cite{Dugan:1984qf}, respectively. ( Usually
$M_2$ is made real by redefining the fields.) From the point of
view of baryogenesis, one might hope that these phases are
large~\cite{Carena:1997ki}. Although the experimental upper bounds
on the electron and neutron electric dipole moments~\cite{Altarev:cf} constrain the phase of $\mu$, $\phi_\mu <
{\cal O}(10^{-2})$~\cite{Nath:dn}, for a typical supersymmetry mass scale
of the order of a few hundred GeV, the phases of the other
parameters mentioned above are practically unconstrained. The
CPV effects that might arise from the trilinear couplings
of the first generation $A_{u,d,e}$ are relatively small as they
are proportional to $m_{u,d,e}$. The same argument holds for the
second generation. Nevertheless, the trilinear couplings of the
third generation $A_{t,b,\tau}$ can lead to significant
CPV effects~\cite{pilaftsis, carenaCP}, especially in top
quark physics~\cite{Atwood:2000tu}.

Recently we studied the effects of CPV in the three possible decay modes
of the MSSM's charged Higgs boson into ordinary particles~\cite{Hplustb, Hplustaunu, HplusWh, Hdecays, Katqproceedings} $H^\pm \to t b$, $H^\pm \to
\nu\tau^\pm$ and $H^\pm \to W^\pm h^0$, where $h^0$ is the lightest
neutral Higgs boson. Loop corrections induced by the MSSM Lagrangian with complex
couplings  lead to non zero decay rate asymmetries between the partial
decay widths of $H^+$ and $H^-$, which is a clear signal
of CPV.

Studying the effects of CPV in the decay $H^\pm \to t b$,
we found that  these effects can be rather large and reach up to 25$\%$~\cite{Hplustb}.
This is mainly due to the contribution of the loop diagrams with stops and sbottoms,
whose couplings are enhanced by the large top quark mass. This motivated our interest
in studying CPV also in the production of $H^\pm$ at LHC
(considered previously in~\cite{Jennifer} and~\cite{jung:lee:song}), where the dominant
production process is the associated production $pp\to H^\pm t +X$, which proceeds
at parton level  through the reaction $bg\to H^\pm t $~\cite{Belyaev:2002eq}.
This process contains the same $H^\pm tb$ vertex and corresponding loop diagrams as the decay
$H^\pm \to tb$, and one would expect that the CPV effects  might be of the same magnitude.
In addition, in the production process there are box graphs that are of the same
order. These contain additional sources of CPV and must also be taken into account.

We assume that the charged Higgs is produced on mass shell and we consider the production and
decay processes separately.
In this paper we first study CPV in $H^\pm$ production  at the
LHC, $pp\to H^\pm t+X$,  through  bottom-gluon fusion in the framework of the MSSM, with running top
an bottom Yukawa couplings. Then we study the CPV asymmetry in the combined process of $H^\pm$
production and decay into $tb$ and $\nu \tau^\pm$, with CPV in both production and decay. We present a detailed
 numerical study for the CPV asymmetry induced by vertex, selfenergy and box corrections in the MSSM.

The paper is organized as follows. In the next section we study the subprocess $b g \to  H^\pm t$ including vertex and
selfenergy loop corrections and obtain analytical expressions for the cross section and the CP-asymmetry at parton level.
In Section 3 we add the parton distribution functions (PDF's) and obtain the CPV asymmetry of the $pp\to H^\pm t +X$ production
process. In Section 4 we obtain the asymmetry in the case of charged Higgs boson production and subsequent decay. Section 5
contains the numerical analysis in the MSSM. We end up with a Conclusion and 3 Appendices, which contain some detailed formulas
needed in the analysis.

\section{The subprocess \boldmath $b g \to t H^\pm$}

We study the following processes connected by charge conjugation
\begin{eqnarray}
b_r (p_b) + g_\mu^\a (p_g) \longrightarrow  t_s (p_t) + H^- (p_{H^-}) \label{pro}\,,~\\
{\bar b}_r (p_{\bar b}) + g_\mu^\a (p_g) \longrightarrow  {\bar t}_s (p_{\bar t}) + H^+ (p_{H^+}) \label{procon}\,,~
\end{eqnarray}
where $r, s $ and $ \a$ are colour indices, $r,s=1,2,3; \a=1,...,8.$ In the kinematics of the processeses we neglect the bottom
mass $m_b$, working in the approximation
$m_b^2/m_t^2 \simeq m_b^2/m_W^2 \simeq 0$. However, we keep $m_b$ non zero in the Yukawa couplings,
where it is multiplied by $\tan\beta$ or $\cot\beta$.

\begin{figure}[h!]
 \begin{center}
 \mbox{\resizebox{!}{3cm}{\includegraphics{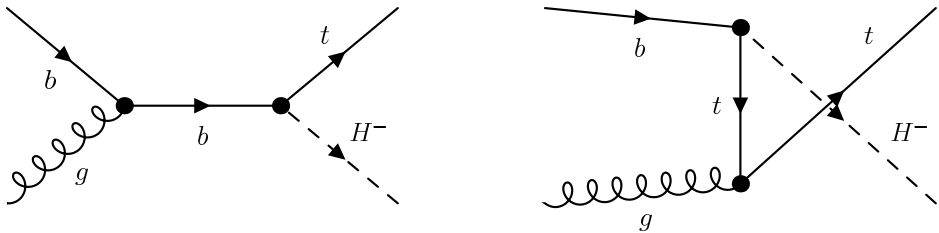}}}
  \end{center}
  \caption {The tree-level graphs of the $b g \to t H^-$ process.
  \label{tree}}
\end{figure}

The tree-level process (\ref{pro}) contains two graphs - with exchange of a bottom quark ($s$-channel) and
with exchange of a top quark ($t$-channel), see
Fig.~\ref{tree}. The Mandelstam
variables are
\begin{equation}
\hat s=(p_b+p_g)^2,\qquad \hat t=(p_t-p_g)^2=(p_b-p_{H^-})^2\,.
\end{equation}

At tree-level there is no difference
between the cross sections of the considered processes
(\ref{pro})~and (\ref{procon}). An asymmetry due to CP non-conservation appears at
one-loop level. There are three types of MSSM loop corrections to both $s$- and
$t$-channels that lead to CPV - corrections to the $H^\pm tb$-vertex, selfenergy corrections on the
$H^\pm$-line and box-type corrections, see Fig.~\ref{contributions}. The first two types, vertex and
selfenergy corrections, are analogous to those in the decay $H^\pm \to tb$. The CPV effects in this
decay were studied in~\cite{Hplustb}.
Our analysis in~\cite{Hplustb} showed that the main contribution to the CPV asymmetry is due to the vertex
diagram with a gluino and the $H^\pm - W^\pm$ selfenergy graph with a $\st \sb$ loop. These contributions are
enhanced by the large top quark mass
and the colour factor of $3$. The contribution of the rest of the graphs with supersymmetric particles is negligible.
We will present analytical expressions for the vertex correction with
$\tilde t \tilde b \tilde g$ and selfenergy correction with $\tilde t \tilde b$ in the loops in the production process.
The expressions for the box-diagram
contributions are rather lengthy and we do not present them analytically, but in the numerical section
they are taken into account.

\subsection{$s$-channel amplitude}

\begin{figure}[h!]
\vspace*{-1cm}
 \begin{center}
  \mbox{\resizebox{!}{11cm}{\includegraphics{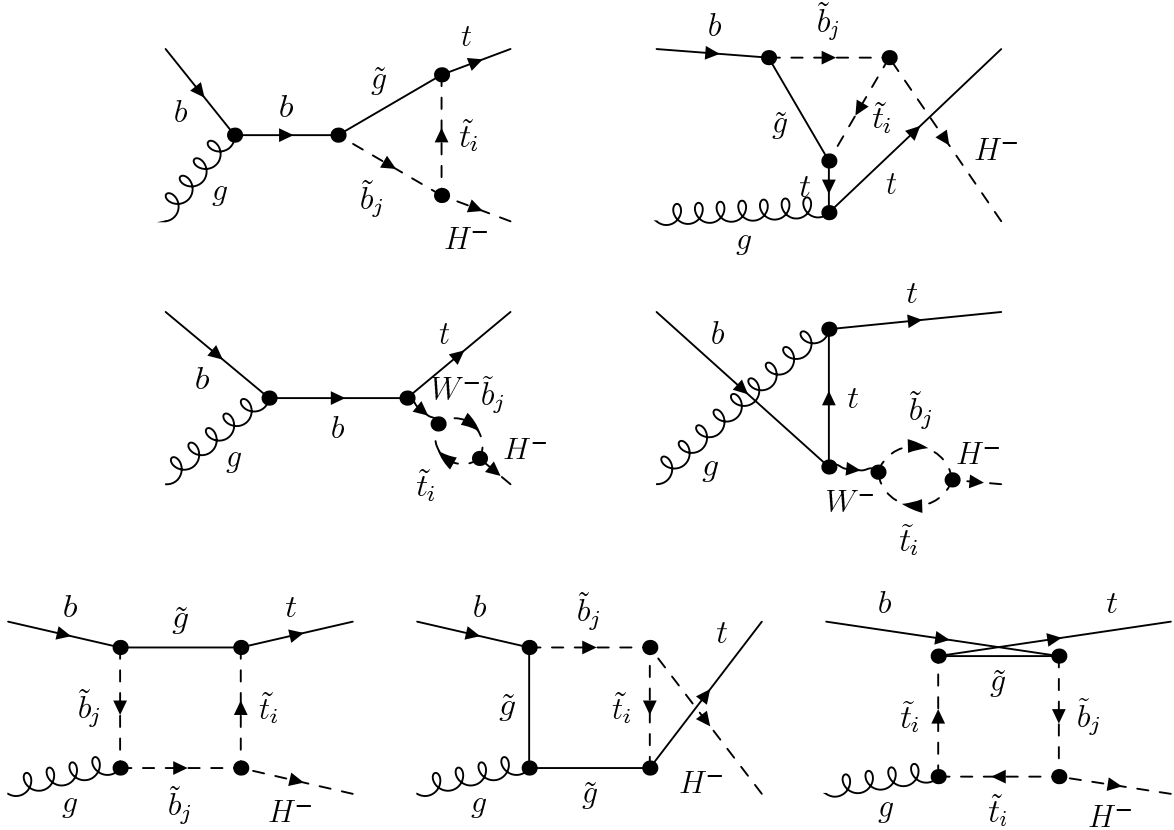}}}
  \end{center}
  \caption{The main sources of CP violation in $b g \to H^\pm t$ production.}
  \label{contributions}
\end{figure}

The matrix element of the graph with bottom exchange on Fig.~\ref{tree} (the $s$-channel),
including the vertex correction
with a gluino in the $H^- tb$-vertex and the $H^- - W^-$ selfenergy graph with a $\st \sb$ loop
(see Fig.~\ref{contributions}) reads\footnote{In this section details are given on the $H^-$-production only.}
\begin{eqnarray}
{\cal M}^s=i \frac{g_s}{\hat s} \bar{u}_s(p_t)\bigg\{[(y_t+\delta \tilde{Y}_t^s) P_L+(y_b+\delta \tilde{Y}_b^s
)P_R]({p
\hspace{-1.8mm} \slash}_b+{p \hspace{-1.8mm} \slash}_g)+ \nonumber \\
 +\hat{s}[f_{RR}^{s, 2}P_L+(f_{LL}^{s, 2}-\tilde f_{LL})P_R]\bigg\}
T_{sr}^{\alpha}\gamma^{\mu}u_r(p_b)\epsilon_{\mu}^{\alpha}(p_g)\,,\label{mats}
\end{eqnarray}
where $y_t$ and $y_b$ are the (real) tree-level couplings and the other terms are induced by the loop corrections.
The principal difference between the $bg \to tH^\pm$ production and the $H^\pm \to tb$ decay, considered
previously in~\cite{Hplustb}, is in the vertex corrections. In the production process one of the quarks
in the $H^\pm tb$-vertex is always off-shell - this is the b quark in the $s$-channel and the t quark in
the $t$-channel. In the  $H^\pm \to tb$ decay all particles are on mass shell. This leads to a different
structure of the matrix elements. The one-loop form factors of the decay repeat the structure of the
tree-level couplings, whereas in the production there are new terms in addition - these terms appear in
the second lines of eq.~(\ref{mats}) for the $s$-channel, and eq.~(\ref{matt}) for the $t$-channel.
For the one-loop form factors in (\ref{mats}) we obtain
\begin{eqnarray}
\delta \tilde{Y}_t^s=m_{\sg} f_{RL}^{s, 0}+m_t (f_{LL}^{s, 1}+\tilde f_{LL})\,,\qquad \delta \tilde{Y}_b^s=
m_{\sg} f_{LR}^{s, 0}+m_t f_{RR}^{s, 1}\,,
\label{eq:dYtbs}
\end{eqnarray}
\begin{eqnarray}
f_{LR}^{s, i}=\frac{2\alpha_s}{3\pi} f_{LR}~C_i^s\,, \qquad f_{LR}={\cal R}^{\st}_{Lm}{\cal R}^{\sb *}_{Rn}e^{i\phi_{\sg}}
(G_4)_{mn}\,,
\end{eqnarray}
\begin{eqnarray}
f_{RL}^{s, i}=\frac{2\alpha_s}{3\pi}  f_{RL}~ C_i^s\,, \qquad  f_{RL}= {\cal R}^{\st}_{Rm}{\cal R}^{\sb *}_{Ln}e^{-i\phi_{\sg}}
(G_4)_{mn}\,,
\end{eqnarray}
\begin{eqnarray}
f_{LL}^{s, i}=\frac{2\alpha_s}{3\pi}f_{LL}~ C_i^s\,,\qquad  f_{LL}={\cal R}^{\st}_{Lm}{\cal R}^{\sb *}_{Ln}(G_4)_{mn}\,,
\end{eqnarray}
\begin{eqnarray}
f_{RR}^{s, i}=\frac{2\alpha_s}{3\pi}f_{RR}~ C_i^s\,, \qquad f_{RR}={\cal R}^{\st}_{Rm}{\cal R}^{\sb *}_{Rn}(G_4)_{mn}\,,
\end{eqnarray}
\begin{eqnarray}
\tilde f_{LL}=-\frac{3 \alpha_w}{8 \pi}\frac{ f_{LL}}{m_W^2}(B_0+2B_1)\,,
\end{eqnarray}
where $\alpha_w = g^2/(4 \pi)$ and the arguments of the Passarino-Veltman (PV) integrals are
\begin{eqnarray}
\qquad \qquad C^s_i=C_i(m_t^2, m_{H^+}^2, \hat s, m_{\sg}^2, m_{\st}^2, m_{\sb}^2), \quad i=0,1,2\, \nonumber  \\
B_j=B_j(m_{H^+}^2,m_{\sb}^2,m_{\st}^2)\,, \quad j=0,1.
\end{eqnarray}
The full expressions for the mixing matrices, the couplings, as well as the definitions of the PV integrals are given in the
Appendices A, B and C, respectively.

\subsection{$t$-channel amplitude}

The matrix element of the graph with a top exchange ($t$-channel) in Fig.~\ref{tree},
including the vertex correction
with a gluino in the $H^- tb$-vertex and the $H^- - W^-$ selfenergy graph with a $\st \sb$
loop (Fig.~\ref{contributions}), reads
\clearpage
\begin{eqnarray}
\hspace*{-1cm}{\cal M}^t & = & i \frac{g_s}{\hat t-m_t^2}
\bar{u}_s(p_t)\epsilon_{\mu}^{\alpha}(p_g)\gamma^{\mu}T_{sr}^{\alpha}
\bigg\{({p\hspace{-1.8mm} \slash}_t-{p \hspace{-1.8mm}
\slash}_g+m_t)[(y_t+\delta \tilde{Y}_t^t) P_L +(y_b+\delta
\tilde{Y}_b^t) P_R] \nonumber \\
&& \hspace*{2cm} + (\hat
t-m_t^2)[(f_{LL}^{t,1}+\tilde f_{LL})P_L+f_{RR}^{t,1}P_R]\bigg\}
u_r(p_b) \,, \label{matt}
\end{eqnarray}
where the one-loop form factors are analogous to those of the $s$-channel
\begin{eqnarray}
\delta \tilde{Y}_t^t=m_{\sg} f_{RL}^{t, 0}+m_t( f_{LL}^{t,1}+ \tilde f_{LL})\,,\qquad \delta \tilde{Y}_b^t=m_{\sg}
f_{LR}^{t, 0}+m_t f_{RR}^{t,1}
\label{eq:dYtbt}
\end{eqnarray}
\begin{eqnarray}
f_{RL}^{t, i}=f_{RL}^{s, i}(C^s_i\rightarrow C^t_i), \quad etc.\,,
\end{eqnarray}
but with different arguments in the PV integrals
\begin{eqnarray}
\qquad \qquad C^t_i=C_i(\hat t, m_{H^+}^2, m_b^2, m_{\sg}^2, m_{\st}^2, m_{\sb}^2), \quad i=0,1,2
\, .
\end{eqnarray}

\subsection{Cross section - parton level}

In general, the differential cross sections for the processes (\ref{pro}) and (\ref{procon}) are
given by
\begin{equation}
d\hat{\sigma}^\pm ={1\over 16\pi \hat{s}^2}{1\over 96}
|{\cal M}^{\pm}|^2~dt\,,
\label{crosssecgeneral}
\end{equation}
where $d\hat{\sigma}^\pm$ are averaged (and summed) over initial (and final) colour and spin of all particles of the
process, we have used $
\Sigma_{s,r=1}^3~\Sigma_{\a=1}^8~
T^\alpha_{sr}(T^{\alpha *}_{sr})=4$, and the $\pm$ signs stand for $H^{\pm}$ production.

We write the matrix elements of the processes (\ref{pro}) and (\ref{procon}),
including the vertex and selfenergy corrections, in the form
\begin{eqnarray}
{\cal M}^{\pm}={\cal M}^{tree, \pm}+{\cal M}^{loop, \pm}.
\end{eqnarray}
Here ${\cal M}^{tree, \pm}={\cal M}_0^{s,\pm}+{\cal M}_0^{t,\pm}$ are the tree-level matrix elements (proportional to $y_{t}$
and $y_b$), and ${\cal M}^{loop, \pm}={\cal M}_1^{s,\pm}+{\cal M}_1^{t,\pm}+{\cal M}_2^{s,\pm}+{\cal M}_2^{t,\pm}$ are the
loop contributions. ${\cal M}_1^{s(t)}$ have the same structure as the tree-level matrix elements
and are proportional to $\delta \ti Y^{s,t}_{t,b}$. ${\cal M}_2^{s(t)}$ contain the additional terms in
(\ref{mats}) and (\ref{matt}).
For the squared matrix elements $|{\cal M}^\pm|^2$, up to terms linear in  $\alpha_s$ and $\alpha_w$, we obtain
 \begin{eqnarray}
 |{\cal M}^\pm|^2=|{\cal M}^{tree, \pm}|^2+2 {\rm Re}\bigg\{({\cal M}^{tree, \pm})^*{\cal M}^{loop, \pm }\bigg\}\,.
 \label{msq}
\end{eqnarray}


Further, we need to sum over the polarizations of the incoming gluon.
At loop level special care must be taken to preserve gauge invariance.
We use the axial gauge:
\begin{eqnarray}
\sum_{\lambda=1}^{2}\epsilon_\mu^{\alpha *}( k,
\lambda)\epsilon_\nu^{\beta}(k, \lambda)= \delta^{\alpha
\beta}\bigg ( -g_{\mu \nu}-\frac{\eta^2 k_\mu k_\nu}{(\eta .
k)^2}+\frac{\eta_\mu k_\nu +\eta_\nu
k_\mu}{\eta.k}\bigg)\,,\label{summation}
\label{eq:polsum}
\end{eqnarray}
where $\eta$ is an arbitrary four-vector that fixes the gauge and fulfills $\eta.\epsilon = 0$ and
\mbox{$\eta. k \ne 0$}. One can see that in this gauge the unphysical longitudinal
degree of freedom manifests itself by an $\eta$-dependent polarization sum
over the two transverse gluon polarizations. The cross section, being a
measurable quantity should be gauge invariant and
therefore the $\eta$-dependence should ultimately cancel.
Using only $-g_{\mu\nu}$ on the right
side of eq.~(\ref{eq:polsum}) is sufficient at tree-level, because the term
resulting from the second and third term on the right side of eq.~(\ref{eq:polsum})
drops out in the squared matrix element of the sum of the $s$- and $t$-channels.
This is not true anymore at one-loop level because
the one-loop factors, e.g. $\delta \tilde Y_{t,b}^s$, eq.~(\ref{eq:dYtbs}), and $\delta \tilde Y_{t,b}^t$,
eq.~(\ref{eq:dYtbt}), are different.\\

For the terms in (\ref{msq}) we obtain
\begin{equation}
\hspace*{-1cm}
{\cal M}^{s*,-}_0{\cal M}^{s,-}_1=-32\pi \alpha_s (\delta \tilde{Y}_t^{s}y_t+\delta \tilde{Y}_b^{s}y_b)[
{\cal X}_1(\hat s,\hat t)-2{\cal U}(\hat s, \hat t) c_\eta]
\label{prod1}\,,
\end{equation}
\begin{equation}
\hspace*{-1cm}
{\cal M}^{t*,-}_0{\cal M}^{s,-}_1=-32\pi \alpha_s (\delta \tilde{Y}_t^{s}y_t+\delta \tilde{Y}_b^{s}y_b)[{\cal X}_{12}(\hat s,
\hat t)+2{\cal U}(\hat s,\hat t) c_\eta +{\cal U}(\hat s,\hat t){\cal V}(\hat s,\hat t)]\,,
\end{equation}
\begin{equation}
\hspace*{-1cm}
{\cal M}^{s*,-}_0{\cal M}^{t,-}_1=-32\pi \alpha_s  (\delta \tilde{Y}_t^{t}y_t+\delta \tilde{Y}_b^{t}y_b)[{\cal X}_{12}(\hat s,
\hat t)+2{\cal U}(\hat s,\hat t) c_\eta +{\cal U}(\hat s,\hat t){\cal V}(\hat s,\hat t)]\,,
\end{equation}
\begin{equation}
\hspace*{-1cm}
{\cal M}^{t*,-}_0{\cal M}^{t,-}_1=-32\pi \alpha_s  (\delta \tilde{Y}_t^{t}y_t+\delta \tilde{Y}_b^{t}y_b)[{\cal X}_{2}(\hat s,
\hat t)-2{\cal U}(\hat s,\hat t) c_\eta -2{\cal U}(\hat s,\hat t){\cal V}(\hat s,\hat t)]\,,
\label{prod4}\end{equation}
\begin{equation}
\hspace*{-1cm}
{\cal M}^{s*,-}_0{\cal M}^{s,-}_2=32\pi \alpha_s m_t [(f_{LL}^{s,2}-\tilde f_{LL})y_t+ f_{RR}^{s,2}y_b] [1+2c_\eta]\,,
\label{prod6}
\end{equation}
\begin{equation}
\hspace*{-1cm}
{\cal M}^{t*,-}_0{\cal M}^{s,-}_2=32\pi \alpha_s m_t [(f_{LL}^{s,2}-\tilde f_{LL})y_t+ f_{RR}^{s,2}y_b][{\cal Y}
(\hat s,\hat t)-2c_\eta - {\cal V}(\hat s,\hat t)]\,,
\end{equation}
\begin{equation}
\hspace*{-1cm}
{\cal M}^{s*,-}_0{\cal M}^{t,-}_2=32\pi \alpha_s m_t [(f_{LL}^{t,1}+\tilde f_{LL})y_t+ f_{RR}^{t,1}y_b][1+2c_\eta]\,,
\end{equation}
\begin{equation}
\hspace*{-1cm}
{\cal M}^{t*,-}_0{\cal M}^{t,-}_2=32\pi \alpha_s  m_t [ (f_{LL}^{t,1}+\tilde f_{LL})y_t+ f_{RR}^{t,1}y_b][{\cal Y}
(\hat s,\hat t)-2c_\eta - {\cal V}(\hat s,\hat t)]
\label{prod8}\,.
\end{equation}
Here $\cal X$, ${\cal U}$, ${\cal V}$ and ${\cal Y}$-functions are given by
\begin{equation}
{\cal X}_1(\hat s,\hat t)=\frac{\hat t-m_t^2}{\hat s}\,,
\end{equation}
\begin{equation}
{\cal X}_{12}(\hat s,\hat t)=\frac{(\hat t-m_{H^+}^2)(\hat s+m_t^2-m_{H^+}^2)-m_t^2 \hat s}{\hat s(\hat t-m_t^2)}\,,
\end{equation}
\begin{equation}
{\cal X}_{2}(\hat s,\hat t)=\frac{\hat s(\hat t-m_t^2)+2 m_t^2(\hat t-m_{H^+}^2)}{(\hat t-m_t^2)^2}\,,
\end{equation}
\begin{equation}
{\cal U}(\hat s,\hat t)=\frac{\hat s+\hat t-m_{H^+}^2}{\hat s}\, , \qquad
{\cal V}(\hat s,\hat t)=\frac{m_{H^+}^2-\hat t}{\hat t-m_t^2}\,,
\end{equation}
\begin{equation}
{\cal Y}(\hat s,\hat t)=\frac{\hat s+m_{H^+}^2-\hat t}{\hat t-m_t^2}\,.
\end{equation}
The terms proportional to the number $c_\eta$ carry the $\eta$-dependence. The calculation is done for two specific choices
of $\eta=p_b; ~ c_\eta=0$ and $\eta=p_b+p_g; ~c_\eta=1$. In the center of mass system,
$\vec p_b+\vec p_g=0$, it is easy to see that in both cases the above conditions $\eta . \epsilon=0$
and $\eta . k \ne 0$ are fulfilled. The sum of the products (\ref{prod1}) --  (\ref{prod8})
is independent on $\eta$ and therefore gauge invariant.

Using eqs.~(\ref{crosssecgeneral}) and (\ref{msq}) for the total parton level cross sections of the processes (\ref{pro})
and (\ref{procon}) we obtain
\begin{eqnarray}
\hat{\sigma}^\pm =\hat{\sigma}^{tree}+{1\over 8\pi \hat{s}^2}{1\over 96}
\int_{t_{min}}^{t_{max}} {\rm Re}\bigg\{({\cal M}^{tree, \pm})^*{\cal M}^{loop, \pm }\bigg\}~dt \,,
\label{totalcspm}
\end{eqnarray}
where the integration limits are given by
\begin{equation}
t_{min, max}={1\over 2}(m_t^2+m_{H^+}^2-\hat{s}\mp \lambda^{1/2}(\hat{s},m_t^2,m_{H^+}^2 ))\,,
\end{equation}
and $\hat{\sigma}^{tree}$ is the tree-level cross section, which
is the same for (\ref{pro}) and (\ref{procon})~\cite{Kidonakis}
\begin{eqnarray}
\hat{\sigma}^{tree}(\hat{s})&=&{\alpha_s \over 96
\hat{s}^3}(y_t^2+y_b^2)\bigg\{\lambda^{1/2}(\hat{s},m_t^2,m_{H^+}^2
)\bigg(7 (m_{H^+}^2- m_t^2)-3\hat{s}\bigg)-\nonumber \\
 &&- 2\bigg(
2(m_{H^+}^2-m_t^2)^2-2(m_{H^+}^2-m_t^2)\hat{s}+\hat{s}^2\bigg)\times
\nonumber \\& &\times
\ln\bigg(\,{\hat{s}-(m_{H^+}^2-m_t^2)-\lambda^{1/2}(\hat{s},m_t^2,m_{H^+}^2
)\over
\hat{s}-(m_{H^+}^2-m_t^2)+\lambda^{1/2}(\hat{s},m_t^2,m_{H^+}^2 )
}\,\bigg)\bigg\}.\label{sigmatree}
\end{eqnarray}

We write the cross sections of the conjugate processes $\hat{\sigma}^\pm$ given with (\ref{totalcspm})
as a sum of CP~invariant and CP violating parts
\begin{equation}
\hspace*{4cm} \hat{\sigma}^\pm=\hat{\sigma}^{inv}\pm \hat{\sigma}^{CP}\,,
\label{cpdiff}
\end{equation}
where the CPV part $ \hat{\sigma}^{CP}$ is given by
\begin{eqnarray}
\hat{\sigma}^{CP}=\frac{\alpha_s}{24 \hat s^2}\bigg\{{\cal A}^{s}\int_{t_{min}}^{t_{max}}({\cal X}_1+{\cal X}_{12}+{\cal U}
{\cal V})~dt\qquad \qquad \qquad \qquad \quad  \nonumber \\
+\int_{t_{min}}^{t_{max}}{\cal A}^{t}({\cal X}_{12}+{\cal X}_{2}-{\cal U}{\cal V})~dt
-\int_{t_{min}}^{t_{max}}({\cal B}^{s}+{\cal B}^{t})(1+{\cal Y}-{\cal V})~dt\bigg\}\,, \label{partdelcp}
\end{eqnarray}
with
\begin{eqnarray}
{\cal A}^{s (t)}=\frac{2 \alpha_s m_{\sg}}{3 \pi} [\,{\rm Im}\,(f_{RL}) y_t+{\rm Im}\,(f_{LR})y_b\,]\,{\rm Im}\,(C_0^{s (t)})+
 \frac{2 \alpha_s m_t}{3 \pi} [\,{\rm Im}\,(f_{LL}) y_t+ \nonumber \\
+{\rm Im}\,(f_{RR}) y_b\,]\,{\rm Im}\,(C_1^{s (t)})-
\frac{3 \alpha_w m_t}{8 \pi}\,{\rm Im}\,( f_{LL}) y_t \frac{\,{\rm Im}\,(B_0+2 B_1)}{m_W^2}\,,
 \label{Astcp}\\
{\cal B}^{s (t)}= \frac{2 \alpha_s m_t}{3 \pi} [\,{\rm Im}\,(f_{LL}) y_t+{\rm Im}\,(f_{RR}) y_b\,]\,{\rm Im}\,(C_{2(1)}^{s (t)})\,,
\qquad \qquad \qquad \qquad
\label{Bstcp}
\end{eqnarray}
and the CP conserving part $ \hat{\sigma}^{inv}$ can be expressed in terms of $ \hat{\sigma}^{CP}$ with exchanging the
imaginary parts of the couplings and the PV integrals with real ones
\begin{eqnarray}
\hat{\sigma}^{inv}=\hat{\sigma}^{tree}- \hat{\sigma}^{CP}\, ( {\rm with}\, {\rm Im} \rightarrow {\rm Re})\,.
\label{partdelinv}
\end{eqnarray}

\subsection{CP violating asymmetry - parton level}

We define the CPV asymmetry at parton level as the difference between the total number of produced
$H^+$ and $H^-$ in bottom-gluon fusion
\begin{equation}
\hat {A}_P^{CP}={\hat{\sigma}(\bar{b} g\rightarrow \bar{t}H^+) -\hat{\sigma}(b g\rightarrow t
H^-)\over \hat{\sigma}(\bar{b} g\rightarrow \bar{t}H^+) +\hat{\sigma}(b g\rightarrow t
H^-)}\,.
\end{equation}
Taking into account (\ref{cpdiff}), we obtain
\begin{eqnarray}
\qquad \qquad \hat {A}_P^{CP}= \frac{\hat{\sigma}^{CP}}{\hat{\sigma}^{inv}}\simeq \frac{\hat{\sigma}^{CP}}{
\hat{\sigma}^{tree}}\,,
\end{eqnarray}
where $\hat{\sigma}^{CP}$ and $\hat{\sigma}^{tree}$ are given by (\ref{partdelcp}) and (\ref{sigmatree}).

\section{The LHC process \boldmath  $pp \to tH^\pm +X$}
%
\subsection{Cross section}

We study charged Higgs boson production associated with top quark production in proton-proton collisions
\begin{equation}
p\, (P_A)+p\, (P_B)\rightarrow t(p_t)+H^\pm(p_{H^\pm}) +X.
\end{equation}
The Mandelstam variable is $S=(P_A+P_B)^2$ ( for LHC $\sqrt{S}=14 $ TeV). We set
$p_b=x_b P_A=\tilde{x}_b P_B$ and $p_g=x_g P_B=\tilde{x}_g P_A$,
where $x_i$ ($\tilde{x_i}$)~is the momentum fraction of the hadron $B (A)$
carried by the parton $i$. Neglecting the proton mass compared to
$\sqrt{S} $ we get $\hat{s}=x_b x_g S=\tilde{x}_b \tilde{x}_g S$. We have

\begin{equation}
\sigma^-(pp\rightarrow t H^-)=2\int_0^1 f_b(x_b)\int_0^1
f_g(x_g)\hat{\sigma}^-(x_b x_g S)\theta (x_b x_g S-S_0)dx_b
dx_g\,,
\end{equation}
\begin{equation}
\sigma^+(pp\rightarrow \bar t H^+)=2\int_0^1 f_{\bar b}(x_{\bar b})\int_0^1
f_g(x_g)\hat{\sigma}^+(x_{\bar b} x_g S)\theta (x_{\bar b} x_g S-S_0)dx_{\bar b}
dx_g\,.
\end{equation}
Here $S_0=(m_t+m_{H^+})^2$ fixes the kinematically allowed energy range,
and $f_b$ and $f_g$ are the PDF's of the bottom and the gluon in the proton. As
$f_{ b}(x_{ b})= f_{\bar b}(x_{\bar b})$, we obtain
\begin{equation}
\sigma^\pm (pp\rightarrow t H^-)=2\int_0^1 f_b(x_b)\int_0^1
f_g(x_g)\hat{\sigma}^\pm (x_b x_g S)\theta (x_b x_g S-S_0)dx_b
dx_g\,.
\label{account}
\end{equation}
The factor 2 in the above expressions counts the two possibilities -- $b\ (g)$ comes from the proton $A\ (B)$ and
${\it vice}$ ${\it versa}$.

\subsection{CP violating asymmetry}

We define the CPV asymmetry at hadron level as the difference
between the total number of produced $H^+$ and $H^-$ in proton-proton collisions
\begin{equation}
A_P^{CP}={\sigma(pp\rightarrow \bar{t}H^+) -\sigma(pp\rightarrow t
H^-)\over \sigma(pp\rightarrow \bar{t}H^+) +\sigma(pp\rightarrow t
H^-)}\,.
\label{APCP}
\end{equation}
Taking into account (\ref{account}) we obtain
\begin{equation}
A_P^{CP}=\frac{\int f_b(x_b) f_g(x_g)(\hat{\sigma}^+ -\hat{\sigma}^-)\theta (x_b x_g S-S_0)dx_b
dx_g}{\int f_b(x_b) f_g(x_g)(\hat{\sigma}^+  +\hat{\sigma}^-)\theta (x_b x_g S-S_0)dx_b
dx_g}\,.
\label{vtoro}
\end{equation}

According to (\ref{cpdiff}), for the CPV asymmetry $A_P^{CP}$
up to terms linear in  $\alpha_s$ and $\alpha_w$, we obtain
\begin{eqnarray}
 \qquad  \qquad  \qquad  \qquad  \qquad A_P^{CP}=\frac{\sigma^{CP}}{\sigma^{tree}}\,,
\end{eqnarray}
where $\sigma^{CP}$ is the CPV part of the cross section
\begin{equation}
\hspace*{-0.5cm}
\sigma^{CP}(pp\rightarrow t H^-)=2\int_0^1 f_b(x_b)\int_0^1
f_g(x_g)\hat{\sigma}^{CP}(x_b x_g S)\theta (x_b x_g S-S_0)dx_b
dx_g\,,
\end{equation}
and $\sigma^{tree}$ is the tree-level
cross section
 \begin{equation}
 \hspace*{-0.5cm}
\sigma^{tree}(pp\rightarrow t H^-)=2\int_0^1 f_b(x_b)\int_0^1
f_g(x_g)\hat{\sigma}^{tree}(x_b x_g S)\theta (x_b x_g S-S_0)dx_b
dx_g\,.
\end{equation}

\section{\boldmath $H^\pm$ production and decay at LHC}

After the charged Higgs is produced in proton-proton collisions it
will be identified through some of its decay modes. Here we
study the combined processes of $H^\pm$ production and decay,
considering $H^\pm$ decays into $t b$, $\nu\tau^\pm$ and $W^\pm h^0$.

\subsection{The subprocess $bg \to tH^\pm \to t t' b ~(t \nu_\tau \tau^\pm;~ t W^\pm h^0)$}

In the narrow width approximation, when the decay width of $H^\pm$  is much smaller than
its mass $m_{H^+}$,  the total cross section for charged Higgs production in $bg \to tH^\pm$,
with a subsequent decay
 $H^\pm \to f$, where $f$ stands for the chosen decay mode
$f=tb;\, \nu\tau^\pm$ and $ W^\pm h^0$, is given by
 \begin{equation}
\hat \sigma^\pm_f=\hat \sigma_P (bg\rightarrow
tH^{\pm})\,\frac{ \Gamma(H^{\pm}\rightarrow f)}{\Gamma_{H^+}}.\label{tot}
\end{equation}
Here, $\hat \sigma_P $ is the total production cross section,
$\Gamma(H^{\pm}\rightarrow f)\equiv \Gamma^\pm_f$ is the corresponding partial decay width of $H^\pm$, and
$\Gamma_{H^+}$ is its total decay width.

We already had the expression for the production parton level
cross section $\hat{\sigma}_P^\pm$ in the form (eq.~(\ref{cpdiff}))
\begin{equation}
{\hat \sigma}_P^\pm
={\hat \sigma}_P^{inv}\pm{\hat \sigma}_P^{CP}\,.
\end{equation}

The considered partial decay widths of $H^\pm$, assuming CPV, were obtained in~\cite{Hplustb, Hplustaunu, HplusWh},
and we write them in the form
\be
\Gamma_f^\pm = \Gamma_0^f \left( \Gamma_f^{inv} \pm \Gamma_f^{CP}\right),\label{decaywidths}
\ee
where $\Gamma^{inv}_f$ and $\Gamma^{CP}_f$ are their CP invariant and
CP violating parts.

For the total cross section of $H^\pm $-production and decay at parton level,
assuming CPV in both  production and decay, we obtain
\begin{equation}
\hat{\sigma}_f^{\pm}=\Gamma^f_0\bigg[\hat{\sigma}_P^{inv}\Gamma_f^{inv}
\pm\bigg(\hat{\sigma}_P^{CP} \Gamma_f^{inv}+
\Gamma_f^{CP}\hat{\sigma}_P^{inv}\bigg)\bigg ]\,.\label{purvo}
\end{equation}

\subsection{CP violating asymmetry - production and decay}

We define the CPV asymmetry in charged Higgs boson production in $pp \to tH^\pm$, with a subsequent decay
 $H^\pm \to f$, assuming CPV in both production and decay, as
\begin{equation}
A^{CP}_{f}={\sigma(pp\rightarrow \bar{t}H^+\to  \bar{t}f)
-\sigma(pp\rightarrow t H^-\to t \bar f)\over
\sigma(pp\rightarrow \bar{t}H^+\to  \bar{t} f)
+\sigma(pp\rightarrow t H^-\to t \bar f)}\,,
\label{AfCP}
\end{equation}
where $f$ stands for the chosen decay mode
$f=t\bar b;\, \nu\tau^+$ and $ W^+ h^0$.

In narrow width approximation, taking into account (\ref{tot}), we get
\begin{equation}
A^{CP}_{f}=\frac {{\sigma}_P(pp\rightarrow
\bar t H^+)\,{\Gamma}(H^+\to f)-
{\sigma}_P(pp\rightarrow
 tH^-)\,{\Gamma}(H^-\to \bar f)}{{\sigma}_P(pp\rightarrow
\bar t H^+)\,{\Gamma}(H^+\to f)+
{\sigma}_P(pp\rightarrow
 tH^-)\,{\Gamma}(H^-\to \bar f)}\, ,\label{4etvurto}
\end{equation}
which leads to~\cite{ Jennifer, jung:lee:song}
\begin{equation}
A^{CP}_f={{\sigma}_P^{inv}\Gamma^{CP}_f+{\sigma}_P^{CP}\Gamma^{inv}_f\over
{\sigma}_P^{inv}\Gamma^{inv}_f}={{\sigma}_P^{CP}\over {\sigma}_P^{inv}}+{ \Gamma^{CP}_f\over
\Gamma^{inv}_f}= A_{P}^{CP}+A^{CP}_{D,f}\,, \label{finalf}
\end{equation}
{\it i.e.} when the decay width of $H^\pm$ is much smaller than
its mass $m_{H^+}$, the total asymmetry $A^{CP}_f$ is an algebraic
 sum of the CPV asymmetry $A^{CP}_P$ in the production,  and the
CPV asymmetry $A^{CP}_{D,f}$ in the decay $f$ of the charged Higgs boson.\footnote{In~\cite{Hptb} the asymmetry
$A^{CP}_D$ in the decay is denoted with $\delta^{CP}$.}

\section{Numerical analysis}
We present numerical results for the charged Higgs rate
asymmetries $A^{CP}_P, A^{CP}_{t b}$ and $A^{CP}_{\nu \tau}$, eqs.~(\ref{APCP}) and (\ref{AfCP}),
in the MSSM. All formulas used in the numerical code are calculated
analytically, except for the box contributions, which are rather
lengthy. Furthermore, all individual one-loop contributions are
checked numerically using the packages FEYNARTS and
FORMCALC~\cite{FeynArts}. We also use LOOPTOOLS, see
again~\cite{FeynArts}, and FF~\cite{FFpackage}. In the numerical
code the Yukawa couplings of the third generation quarks ($h_t$,
$h_b$) are taken to be running~\cite{Hplustb}, at the scale $Q =
m_{H^+} + m_t$. For the evaluation of the PDF's of the bottom quark and the gluon, $f_b$ and $f_g$,
we use CTEQ6L~\cite{PDFs}, with leading order PDF's and next-to-leading order $\alpha_s$, at the
same scale $Q$. We assume the grand unified theory relation between $M_1$ and $M_2$, so that
the phase of $M_1=0$. Our numerical study shows that the
contributions of the loop diagrams with chargino, neutralino, stau
and sneutrino to the considered CPV asymmetries are
negligible, and besides one exception we show only the contributions from
diagrams with $\st \sb$ and $ \sg$. We start from the following
reference scenario:
\begin{eqnarray}
\hspace*{-1.5cm} && \tan \beta = 5\,,\quad M_2=300~{\rm
GeV}\,,\quad m_{\tilde g} =727~{\rm GeV}\,, \quad M_{\tilde U} =
M_{\tilde Q} = M_{\tilde D} = 350~{\rm GeV}\,,\nonumber\\
\hspace*{-1.5cm} && \mu=-700~{\rm GeV}\,,\quad |A_t| = |A_b|
=700~{\rm GeV}\, ,\quad \phi_{A_t}={\pi \over 2}\, ,\quad
\phi_{A_b}=\phi_{\mu} = \phi_{3} = 0\, . \label{parameters}
\end{eqnarray}
The relevant masses of the sparticles for this choice of
parameters, and also for $\tan \beta = 30$ are shown in
Table~\ref{table:1}.

As we will see, in such a scenario the effects of CPV are substantial. It is not one of the commonly
used minimal supergravity or constrained MSSM scenarios~\cite{constraints}, for which most studies have been done. There exist experimental constraints
from $b \to s \gamma$, relic density, etc.. In principle, there are enough free parameters in the general complex MSSM to be compatible
with all data.  
\newcommand{\ix}{@{\hspace{3mm}}}
\begin{table}[h!]
\begin{center}
\begin{tabular}{|c||c\ix c\ix c\ix c|c\ix c|c\ix c|c\ix c|c\ix c|c|}
\hline
   $\tan\beta$
  & $m_{{\tilde \chi}^0_1}$ & $m_{{\tilde \chi}^0_2}$ & $m_{{\tilde \chi}^0_3}$ & $m_{{\tilde \chi}^0_4}$ &
   $m_{{\tilde \chi}^+_1}$ & $m_{{\tilde \chi}^+_2}$ &
   $m_{{\tilde t}_1}$ &  $m_{{\tilde t}_2}$  &  $m_{{\tilde b}_1}$  &  $m_{{\tilde b}_2}$  &
   $m_{{\tilde \tau}_1}$ & $m_{{\tilde \tau}_2}$ & $m_{{\tilde \nu}}$ \\
\hline
 5 & 142 & 300 & 706 & 706 & 300 & 709 & 166 & 522 & 327 & 377 & 344 & 362 & 344 \\
 30 & 141 & 296 & 705 & 709 & 296 & 711 & 172 & 519 & 183 & 464 & 295 & 402 & 344\\
\hline
\end{tabular}
\end{center}
\caption{Masses of the sparticles (in GeV) for the parameter set
(\ref{parameters}). \label{table:1}}
\end{table}

\begin{figure}[h!]
\begin{center}
\mybox{\resizebox{85mm}{!}{\includegraphics{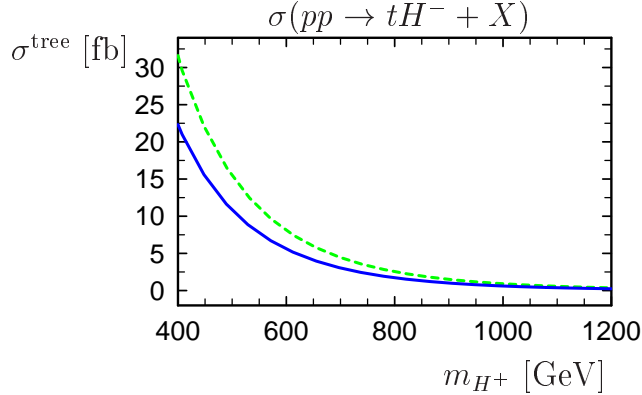}}}
\end{center}
\caption {The hadron tree-level cross section for the chosen set
of parameters (\ref{parameters}) as a function of $m_{H^+}$. The
green dashed line corresponds to the result using on-shell and the
solid blue line to that using running $h_t$ and $h_b$.}
\label{figsigtree}
\end{figure}

In Fig.~\ref{figsigtree} the tree-level cross section $\sigma(p p
\to t H^- + X)$ is shown as a function of $m_{H^+}$, based only on
the parton process $g b \to t H^-$, with on-shell and with running $h_t$ and $h_b$. Taking running $h_t$ and
$h_b$ reduces $\sigma^{tree}$ by about $\sim$~30\%. For $m_{H^+}
\gsim $~1000~GeV the cross section drops below 1~fb.

\subsection{Production asymmetry}

As expected, the CPV asymmetry in the production due to the
loop corrections with $\st \sb $ and $\sg $ is of the same order
of magnitude as in the case of the decay $H^\pm \to
t b$~\cite{Hplustb}, and can go up to $\sim 20$\%. Moreover, the
contributions of the box graphs are significant and can be dominant for
relatively small $m_{H^+}$.

\begin{figure}[h!]
\begin{center}
\mybox{\resizebox{85mm}{!}{\includegraphics{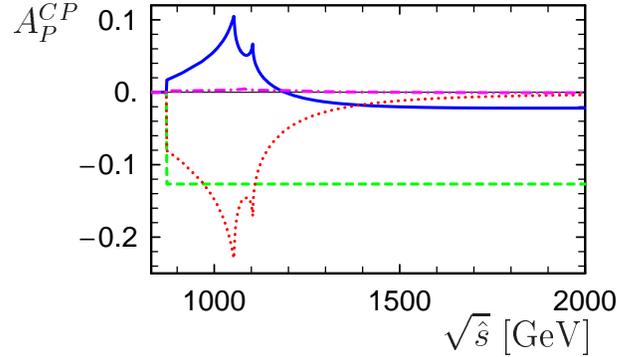}}}
\end{center}
\caption {The contributions to the asymmetry $A_P^{CP}$ at parton
level for the chosen set of parameters (\ref{parameters}) as a
function of $\sqrt{\hat s}$, $m_{H^+} =$~700~GeV. The red dotted
line corresponds to box graphs with a gluino, the solid blue one to
the vertex graph with a gluino, the green dashed one to the $W^\pm -
H^\pm$ selfenergy graph with a $\tilde t \tilde b$ loop, and the
pink dash-dotted to the box and vertex graphs with
$\tilde\chi^{0/+}$.} \label{figshat}
\end{figure}

In Fig.~\ref{figshat} the dependence of the different contributions
on $\sqrt{\hat s}$ of the underlying parton process $g b \to t
H^-$ is shown for $m_{H^+} =$~700~GeV. The kinematical threshold
is at $\sqrt{\hat s} = 871.4$~GeV. First the box contribution is
the biggest one with a maximum at $\sim - 23$~\%. Then it drops down
asymptotically to zero at $\sqrt{\hat s} \gsim $~1700~GeV. The
vertex contribution has a similar shape with about half of the size
of the box contribution. But for $\sqrt{\hat s} > $~1500~GeV it
becomes constant being of $-2$\%. The selfenergy contribution is
independent of $\sqrt{\hat s}$, about  $-12$~\%. We show it only for
completeness. Here also the box and vertex contribution with
$\tilde\chi^{0/+}$, denoted by the pink dash-dotted line, are
shown. This contribution is always below 0.5\%, and therefore we will
not show it anymore in the figures. The two spikes in the box and in
the vertex contributions denote the two thresholds $\sqrt{\hat s}
= m_{\tilde g} + m_{\tilde b_{1,2}}$, see Table~\ref{table:1}.

\begin{figure}[h!]
\begin{center}
\mybox{\resizebox{85mm}{!}{\includegraphics{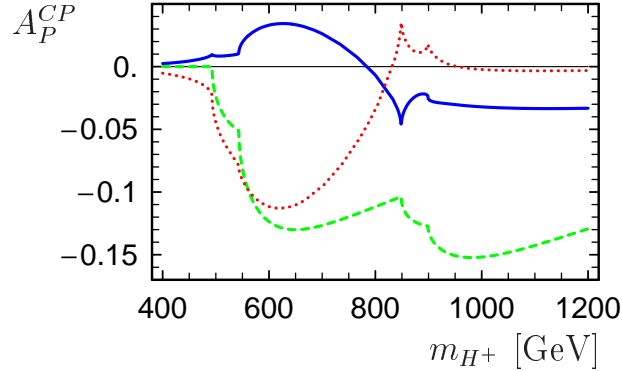}}}
\end{center}
\caption { The contributions to the asymmetry $A_P^{CP}$ at hadron
level for the chosen set of parameters (\ref{parameters}) as a
function of $m_{H^+}$. The red dotted line corresponds to box
graphs with a gluino, the solid blue one to the vertex graph with
a gluino, and the green dashed one to the $W^\pm - H^\pm$ selfenergy
graph with a $\tilde t \tilde b$ loop.} \label{figPHp}
\end{figure}

The contributions of the vertex, selfenergy and box graphs with
$\st \sb $ and $\sg$ to the asymmetry $A_P^{CP}$ at hadron level
as functions of $m_{H^+}$ are shown on Fig.~\ref{figPHp}. The
large effect seen on the figure is mainly due to the phase of
$A_t$, and the asymmetry reaches its maximum for a maximal phase
$\phi_{A_t}=\pi /2$. The phase of $A_b$ does not have a big
influence on the asymmetry and therefore we usually set it
zero. The four kinks in all three lines denote the thresholds
$m_{H^+} = m_{\tilde t_i} + m_{\tilde b_j}\, , i, j = 1,2$, see
again Table~\ref{table:1}.

\begin{figure}[h!]
\begin{center}
\mybox{\resizebox{85mm}{!}{\includegraphics{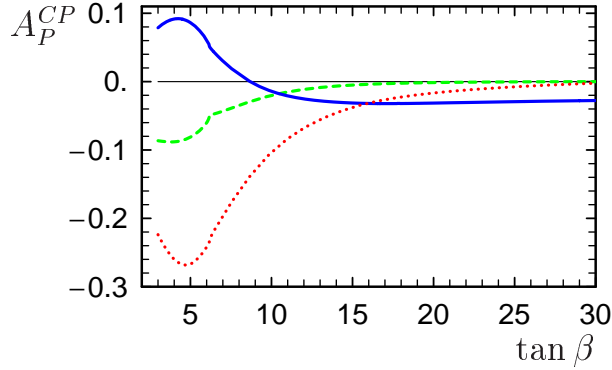}}}
\end{center}
\caption {The contributions to the asymmetry $A_P^{CP}$ at hadron
level for the chosen set of parameters (\ref{parameters}), but
$m_{\tilde g} =$~450~GeV, as a function of $\tan\beta$, $m_{H^+}
=$~550~GeV. The red dotted line corresponds to box graphs with
a gluino, the solid blue one to the vertex graph with a gluino, and
the green dashed one to the $W^\pm - H^\pm$ selfenergy graph with
a $\tilde t \tilde b$ loop.} \label{figTB}
\end{figure}

The asymmetry $A_P^{CP}$ reaches its maximum value at $\tan \beta
=5$ and falls down quickly with increasing $\tan \beta$. This
dependence for $m_{H^+}=550$ GeV is shown on Fig.~\ref{figTB}.

\begin{figure}[h!]
\begin{center}
\mybox{\resizebox{85mm}{!}{\includegraphics{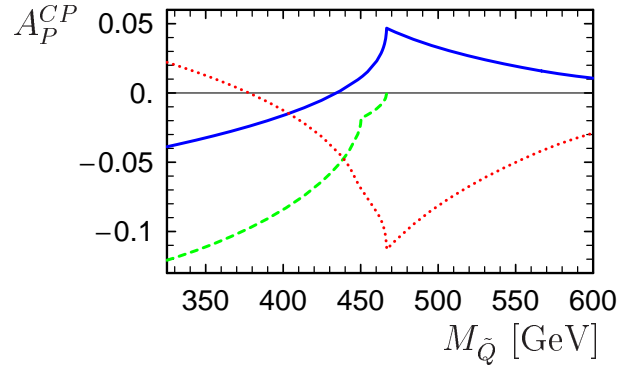}}}
\end{center}
\caption {The contributions to the asymmetry $A_P^{CP}$ at hadron
level for the chosen set of parameters (\ref{parameters}), but
$m_{\tilde g} =$~435~GeV, and $m_{H^+} = $~800~GeV, as a function
of $M_{\tilde Q}$. The red dotted line corresponds to box graphs
with a gluino, the solid blue one to the vertex graph with a gluino,
and the green dashed one to the $W^\pm - H^\pm$ selfenergy graph
with a $\tilde t \tilde b$ loop.} \label{figMSQ3}
\end{figure}

In Fig.~\ref{figMSQ3} we present the dependence of $A_P^{CP}$ as a
function of $M_{\tilde Q} (= M_{\tilde U} = M_{\tilde D})$. The
selfenergy contribution is first the biggest one, but it goes down
to zero at $M_{\tilde Q} \sim 467$~GeV because then the decay
channel $H^+ \to \tilde t_1 \tilde b_1$ closes. The kink at
$M_{\tilde Q} \sim 450$~GeV denotes the threshold  $H^+ \to \tilde
t_1 \tilde b_2$. The box contribution has its maximum of
$\sim$~12\% at the threshold of $H^+ \to \tilde t_1 \tilde b_1$

\begin{figure}[h!]
\begin{center}
\mybox{\resizebox{85mm}{!}{\includegraphics{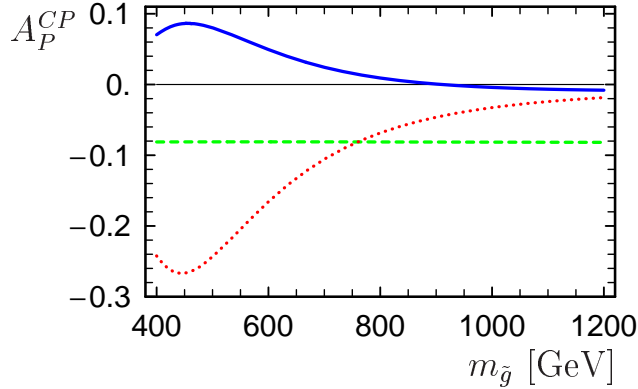}}}
\end{center}
\caption {The contributions to the asymmetry $A_P^{CP}$ at hadron
level for the chosen set of parameters (\ref{parameters}) as a
function of $m_{\tilde g}$, $m_{H^+} = 550$GeV. The red dotted
line corresponds to box graphs with a gluino, the solid blue one to
the vertex graph with a gluino, and the green dashed one to the
$W^\pm - H^\pm$ selfenergy graph with a $\tilde t \tilde b$ loop.}
\label{figPsg}
\end{figure}

In Fig.~\ref{figPsg} the dependence of the three leading
contributions to $A_P^{CP}$ as a function of $m_{\tilde g}$ is
shown for $m_{H^+} = 550$ GeV. Of course, the selfenergy
contribution is independent of the gluino mass, being about $- 8$~\%.
The vertex contribution has a maximum and the box contribution a minimum at
$m_{\tilde g} \sim 425$~GeV, and then their absolute values decrease. The box contribution is
at the minimum about $-27$~\%.

\subsection{Production and decay asymmetry}

\begin{figure}[h!]
\begin{center}
\mybox{\resizebox{85mm}{!}{\includegraphics{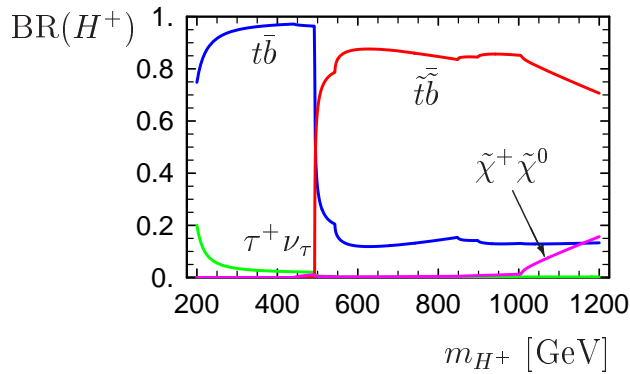}}}
\end{center}
\caption {The tree-level branching ratios of $H^+$ for the chosen
set of parameters (\ref{parameters}) as a function of $m_{H^+}$.}
\label{figBR}
\end{figure}

First we want to add a few remarks on the branching ratios (BR) of
the relevant decays. Fig.~\ref{figBR} shows the tree-level BRs of
$H^+$ as functions of $m_{H^+}$. For small $m_{H^+}$, below the
$\st \sb$ threshold, the dominant decay mode is $H^\pm \to t b$,
with BR $ \approx 1$, while the BR of $H^\pm \to \nu \tau^\pm$ is
of the order of a few percent, decreasing with increasing
$m_{H^+}$. When the $H^\pm \to \tilde t \tilde b$ channels are
kinematically allowed, they start to dominate~\cite{stopsbotBR},
and the BR of $H^\pm \to \nu \tau^\pm$ to a good approximation becomes zero.
However, the BR of  $H^\pm \to t b$ remains stable of the order of
$15-20$~\%. The BR of $H^\pm \to W^\pm h^0$ reaches a few percent
for small $\tan\beta$ in a relatively narrow range of
$m_{H^+}$~\cite{HplusWh}. In the considered range of parameters
this decay is very much suppressed and we do not investigate it
numerically.

\begin{figure}[h!]
\begin{center}
\mybox{\resizebox{85mm}{!}{\includegraphics{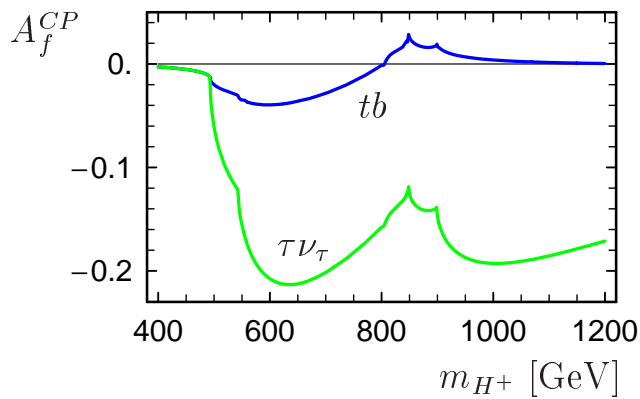}}}
\end{center}
\caption { The total asymmetry $A^{CP}$ at hadron level for the
chosen set of parameters (\ref{parameters}) as a function of
$m_{H^+}$. The blue line corresponds to the case where $H^\pm$
decays to $t b$, and the green one to $H^\pm$ decay to $\tau
\nu_\tau$.} \label{figHp}
\end{figure}

In Fig.~\ref{figHp} we show the total production and decay
asymmetry $A^{CP}_f$ at hadron level, for $f= t b$ and $f=
\nu\tau^\pm$. Though for $H^\pm \to \nu \tau^\pm$ it can go up to
$\sim 20$\% for $m_{H^+}\approx 650$ GeV, the BR of this decay in
this range of $H^+$ masses is too small and observation at LHC is
impossible.

As we have shown analytically, the total asymmetry in the production
and decay is approximately the algebraic sum of the
asymmetry in the production $A^{CP}_P$, and the asymmetry in the
decay $A^{CP}_D$. One would think that the total
CPV asymmetry will be large. Moreover, the
CP-asymmetry in the decay alone is large~\cite{Hplustb}.

\begin{figure}[h!]
 \begin{center}
 \begin{picture}(160,122)
\put(-90,0){
\mbox{\resizebox{!}{4cm}{\includegraphics{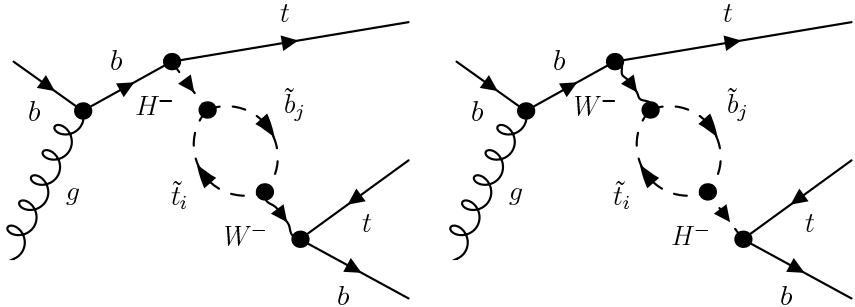}}}}
\hfil
 \end{picture}
 \end{center}
 \caption {The selfenergy contributions with $\st \sb$ in the loop
 to the s-channel of the considered process,
 from the production and from the decay.}
 \label{cancellation}
\end{figure}

\begin{figure}[h!]
\begin{center}
\mybox{\resizebox{85mm}{!}{\includegraphics{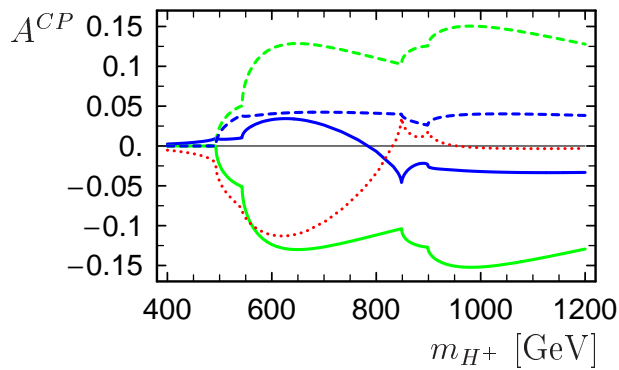}}}
\end{center}
\caption {The contributions to the total asymmetry $A_{t b}^{CP}$
at hadron level for the chosen set of parameters
(\ref{parameters}) as a function of $m_{H^+}$, with the same
parameter set as used in Fig.~\ref{figPHp}. The dotted (red) line
corresponds to box graphs with a gluino, the axially symmetric solid and
dashed (green) lines correspond to
the $W^\pm - H^\pm$ selfenergy graph with a $\tilde t \tilde b$ loop in
the production and the decay, respectively, and the other two solid and
dashed (blue) lines correspond to the vertex graph with a gluino,
again in the production and the decay, respectively.} \label{figcancel}
\end{figure}

Let us consider the subsequent decay $H^\pm \to tb$ decay. In
this case the contributions coming from the selfenergy graph with
$\st \sb$ in the loop in the production and in the decay exactly
cancel. This cancellation occurs in general for any possible $H^\pm$
selfenergy loop contribution to the $s$- or $t$-channel. It can be
easily shown by writing down the matrix element for the whole
three particle final state process. As an illustrative example,
let us consider the contribution of both selfenergy graphs with
$\st \sb$ from the production and from the decay, to the
$s$-channel of our process, see Fig.~\ref{cancellation}. The
matrix element, representing the sum of the two graphs in
Fig.~\ref{cancellation} reads
\begin{eqnarray}
\hspace*{-1.5cm} && {\cal M}^{3, s} =  -\frac{i g_s}{\hat
s}\frac{\,\,g^2}{2}\bar u (p_b) (y_t P_R+y_b P_L)u(-p_t) \bigg
[(G_4)_{ij}{\cal R}^{\st}_{Li}{\cal R}^{\sb
*}_{Lj}+(G_4)^*_{ij}{\cal R}^{\st *}_{Li}{\cal R}^{\sb}_{Lj}\bigg]
\times \nonumber \\ \hspace*{-1.5cm} && \int_{q}
\frac{(p_{\st_i}+p_{\sb_j})^\mu (g_{\mu
\nu}-\frac{p_{W_\mu}p_{W_\nu}}{m_{W}^2})}{(p_{H^+}^2-m_{H^+}^2)(p_{\st_i}^2-
m_{\st_i}^2)(p_{\sb_j}^2-m_{\sb_j}^2)(p_{W}^2-m_{W}^2)}\bar
u_s(p_t)\gamma^\nu P_L {p \hspace{-1.8mm} \slash}_b T_{sr}^\alpha
\gamma^\lambda u_r (p_b)\epsilon_\lambda^\alpha (p_g)\,,\nonumber
\\ \hspace*{-1.5cm} && \label{cancel}
\end{eqnarray}
with $\int_q = \int {\rm d}^D q/(2 \pi)^D$, $p_{\st_i} = p_{H^+} +
q$ and $p_{\sb_j} = q$. It is clearly seen that (\ref{cancel})
contains the sum of the couplings and their complex conjugate ones
as a common factor. The expression in the box brackets at the end
of the first row can be written as
\be
(G_4)_{ij}{\cal R}^{\st}_{Li}{\cal R}^{\sb
*}_{Lj}+(G_4)^*_{ij}{\cal R}^{\st *}_{Li}{\cal R}^{\sb}_{Lj}=2
{\rm Re}[(G_4)_{ij} {\cal R}^{\st}_{Li}{\cal R}^{\sb }_{Lj}] \,,
\ee {\it i.e.} in this case the imaginary part of the couplings
cancels. As the presence of a non zero imaginary part of the
couplings is necessary for having CPV, it is clear that
in this case the contribution of the selfenergy graphs on
Fig.~\ref{cancellation} to the CPV asymmetry is exactly
zero. Our numerical study shows that the
contributions of the vertex graphs from the production and from the decay also
partially cancel with the box diagrams contribution. However, as the box graphs do not have
an analogue in the decay, their
contribution remains the leading one, see Fig.~\ref{figcancel}.

\begin{figure}[h!]
\begin{center}
\mybox{\resizebox{85mm}{!}{\includegraphics{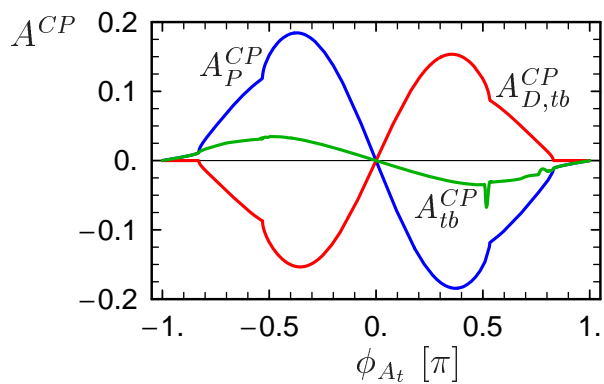}}}
\end{center}
\caption {The asymmetry $A^{CP}$ at hadron level for the
production and the decay into $t b$ only, and the total one,
for the chosen set of parameters (\ref{parameters}) as a function of
$\phi_{A_t}$.} \label{figphiAt}
\end{figure}

Fig.~\ref{figphiAt} shows the asymmetry in the production,
$A^{CP}_{P}$, in the decay of $H^\pm \to t b$, $A^{CP}_{D, t b}$,
and the combined one, $A^{CP}_{tb} = A^{CP}_{P} + A^{CP}_{D, t b}$, as
a function of the phase $\phi_{A_t}$. All three curves are
symmetric for $\phi_{A_t} \to -\phi_{A_t}$ because $\phi_{\mu} =
0$. $A^{CP}_{P}$ and $A^{CP}_{D, t b}$ have negative relative
signs with a maximium/minimum at $|\phi_{A_t}| \lsim 0.45 \pi$,
with $|A^{CP}_{P}| \sim$~20\% and $|A^{CP}_{D, t b}| \sim$~16\%
there. Due to this cancellation the resulting asymmetry
$A^{CP}_{t b}$ is reduced to $\lsim 4$\%.

In order to avoid the cancellation, we can study the mass range of
$H^\pm$, before the $\st_1 \sb_1$ channel opens. In this case,
having in mind our results in~\cite{Hplustb}, the CP effects in
the decay will be negligible and CPV will arise mainly in the
production process due to the vertex and box contributions with
$\tilde g$ in the loops. For instance, for $m_{H^+} =$~400~GeV and
$m_{\tilde g} =$~450~GeV we get $A^{CP}_{t b} = - 3$\%.

\begin{figure}[h!]
\begin{center}
\mybox{\resizebox{85mm}{!}{\includegraphics{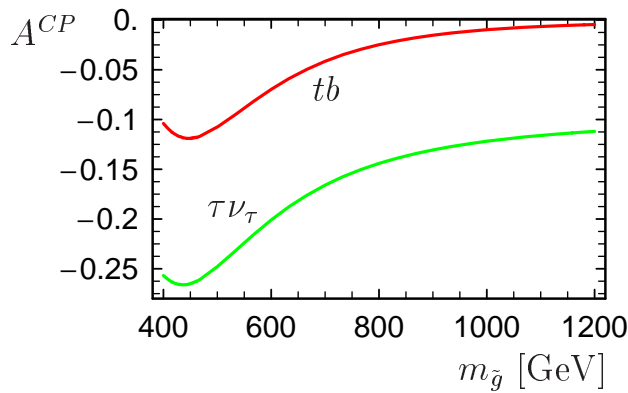}}}
\end{center}
\caption {The total asymmetry $A^{CP}$ at hadron level for the
chosen set of parameters (\ref{parameters}) as a function of
$m_{\tilde g}$, $m_{H^+} = 550$~GeV. The (red) line denoted by $tb$
corresponds to the case where $H^\pm$ decays to $t b$, and the (green)
one denoted by $\tau \nu_\tau$ to
$H^\pm$ decay to $\tau \nu_\tau$.} \label{figsg}
\end{figure}

In Fig.~\ref{figsg} the total asymmetries $A^{CP}_{t b}$ and
$A^{CP}_{\nu \tau}$ as functions of $m_{\tilde g}$ are shown for
$m_{H^+} =$~550~GeV. Both
asymmetries are negative. They have their largest values at $m_{\tilde g} \sim 425$~GeV,
with $A^{CP}_{\nu \tau} \sim  -27$~\%, $A^{CP}_{tb} \sim  -12$~\%, and
then they decrease to zero.

\begin{figure}[h!]
\begin{center}
\mybox{\resizebox{85mm}{!}{\includegraphics{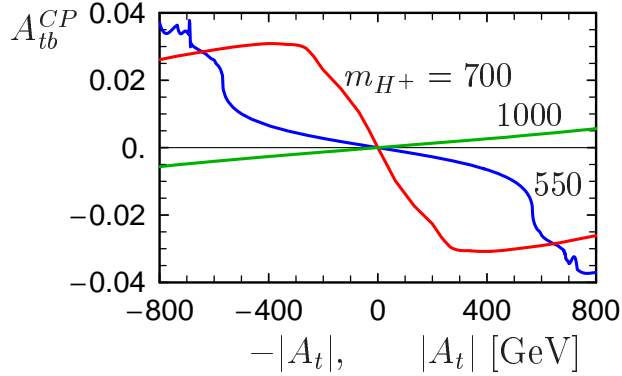}}}
\end{center}
\caption {The total asymmetry $A_{tb}^{CP}$ at hadron level for the
chosen set of parameters (\ref{parameters}) as a function of
$|A_t|$, for three values of $m_{H^+}$ (in GeV).} \label{figabsAt}
\end{figure}

In Fig.~\ref{figabsAt} the dependence of $A^{CP}_{tb}$ on the
absolute value of $A_t$ is shown for three different values of $m_{H^+}$.

\begin{figure}[h!]
\begin{center}
\mybox{\resizebox{85mm}{!}{\includegraphics{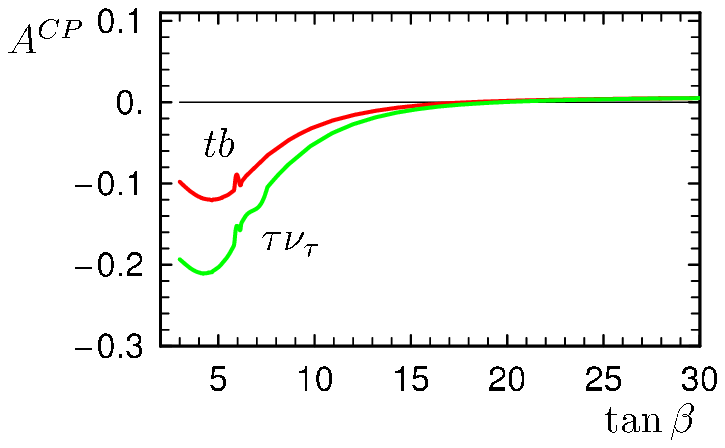}}}
\end{center}
\caption {The total asymmetry $A^{CP}$ at hadron level for the
chosen set of parameters (\ref{parameters}) as a function of $\tan
\beta $, $\msg=450$~GeV, $m_{H^+} = 550$~GeV. The (red) line
denoted by $tb$ corresponds to the case where $H^\pm$ decays to $t b$, and the
(green) one denoted by $\tau \nu_\tau$ corresponds to $H^\pm$ decay to
$\tau \nu_\tau$.} \label{figtbppd}
\end{figure}

As already mentioned in the introduction, the phase of $\mu$ is
strongly constrained by the measurements of the EDMs.
Nevertheless, we also studied the dependence of $A^{CP}$ on
$\phi_\mu$. Using $\mu = 700\, e^{- i {\pi \over 3}}$~GeV instead
of $\mu = -700$~GeV, we get for $m_{\tilde g} = 450$~GeV in
Fig.~\ref{figsg} the asymmetry $A_{t b}^{CP} \sim - 22$\%.

The $\tan \beta$ dependence is shown on Fig.~\ref{figtbppd}, for
 $f=tb$ and $f=\tau \nu_{\tau}$. In both considered cases the asymmetry $A^{CP}$
has its maximum at $\tan \beta \sim 5$, with $A_{tb}^{CP} \sim 12 \%$ and $A_{\nu \tau}^{CP} \sim 21 \%$. It approximately vanishes for $\tan \beta \gsim 15$.

We have compared our results with those in \cite{Jennifer,jung:lee:song}.
Our numerical results are in good agreement with~\cite{Jennifer},
but we disagree analytically and numerically
with~\cite{jung:lee:song}, where in addition the box contributions
are missing.

The production rate of $H^+$ at the LHC for $m_{H^+} = $~550~GeV
and $\tan\beta = 5$ is $\sim 15$~fb (including a K-factor from QCD
of 1.5, see \cite{Belyaev:2002eq}), and the BR($H^+ \to t b$)
$\sim$~20\%. For an integrated luminosity of 200~fb$^{-1}$ we get
$N \sim 600$ and $\sqrt{N} = 25$. For $m_{\tilde g} = 450$ we get
$A^{CP}_{t b}\sim 0.12$ and therefore the statistical significance
$\sqrt{N} A^{CP}_{t b} \sim 4$. But because of the large
background, the actual signal production rate will be most likely reduced
and the statistical significance might be too low for a clear
observation in $H^\pm t$ production in the first stage of LHC. However, at
SLHC with a design luminosity bigger by a factor of $\sim$~10, such
a measurement would be worth of being performed.

\section{Conclusions}
The MSSM with complex parameters in particular with $A_t$ complex,
gives rise to CP~violation in the production of $H^\pm$,
$p p \to H^\pm t + X$, and in the decays of $H^\pm$ to $t b,\, \nu_\tau \tau^\pm$
and $W^\pm h^0$ at one-loop level. We have calculated the
corresponding asymmetries between the $H^+$ and $H^-$ rates both
in the production and in the decays. A few improvements have been
made with respect to previous calculations.

We have performed a detailed numerical analysis studying the
dependence on the important parameters. A peculiarity is that in the case of
$p p \to H^\pm t + X$ with $H^\pm \to t b$, the contributions
coming from the selfenergy graph with $\tilde t \tilde b$ in the
loop in the production and in the decay exactly cancel.
Nevertheless, the asymmetry can go up to $\sim 12$\% for $m_{H^+}
\simeq 550$~GeV, mainly due to the box graphs with gluino in the production
process. A measurement of the CP~violating asymmetries at LHC can give
important information of the parameters of the MSSM, especially on
$A_t$ and its phase.

\section*{Acknowledgments}
%

The authors acknowledge support from EU under the MRTN-CT-2006-035505 network
programme. This work is supported by the "Fonds zur F\"orderung der
wissenschaftlichen Forschung" of Austria, project No. P18959-N16. The work of
E.~C. and E.~G. is partially supported by the
Bulgarian National Science Foundation, grant 288/2008.
%

\begin{appendix}

\section{Masses and mixing matrices}

The mass matrix of the stops in the basis $(\tilde t_L,\,\tilde t_R)$ reads
\begin{equation}
\hspace*{-1cm}
  {\cal M}_{\tilde t}^2 = {\small \left(
  \begin{array}{cc}
    M_{\ti Q}^2 + m_Z^2\cos 2\beta(\frac{1}{2}-\frac{2}{3}\sin^2\theta_W)
    + m_t^2 & (A_t^*-\mu\cot\beta)\,m_t \\
    (A_t-\mu^*\cot\beta)\,m_t
    & M_{\ti U}^2 + \frac{2}{3}m_Z^2\cos 2\beta\sin^2\theta_W + m_t^2
  \end{array}\right)}\,.
\end{equation}
${\cal M}_{\tilde t}^2$ is diagonalized by the rotation matrix
${\cal R}^{\,\st}$ such that
${\cal R}^{\,\st\,\dagger}\,{\cal M}_{\tilde t}^2\, {\cal R}^{\,\st} =
 {\rm diag}(m_{\tilde t_1}^2,\,m_{\tilde t_2}^2)$ and
${\scriptsize \Big(\!\!\begin{array}{cc} \st_L \\ \st_R \end{array}\!\!\Big)} = {\cal R}^{\tilde t} \,
 {\scriptsize \Big(\!\!\begin{array}{cc} \st_1 \\ \st_2 \end{array}\!\!\Big)}$.

We have
\begin{equation}
  {\cal R}^{\,\tilde{t}} =
    \left(\begin{array}{rr}
      {\cal R}^{\,\tilde{t}}_{L1} & {\cal R}^{\,\tilde{t}}_{L2} \\
      {\cal R}^{\,\tilde{t}}_{R1} & {\cal R}^{\,\tilde{t}}_{R2}
    \end{array}\right)
  = \left(\begin{array}{rr}
      e^{\frac{i}{2} \varphi_{\tilde{t}}} \cos\theta_{\tilde{t}}
    & -e^{\frac{i}{2} \varphi_{\tilde{t}}} \sin\theta_{\tilde{t}}
  \\ e^{-\frac{i}{2} \varphi_{\tilde{t}}} \sin\theta_{\tilde{t}}
    & e^{-\frac{i}{2} \varphi_{\tilde{t}}} \cos\theta_{\tilde{t}}
    \end{array}\right) \;. \label{stopsmatrix}
\end{equation}
Analogously, the mass matrix of the sbottoms
in the basis $(\tilde b_L,\,\tilde b_R)$
\begin{equation}
\hspace*{-1cm}
  {\cal M}_{\sb}^2 = {\small \left(
  \begin{array}{cc}
    M_{\ti Q}^2 - m_Z^2\cos 2\beta(\frac{1}{2}-\frac{1}{3}\sin^2\theta_W)
    + m_b^2 & (A_b^*-\mu\tan\beta)\,m_b \\
    (A_b-\mu^*\tan\beta)\,m_b
    & M_{\ti D}^2 - \frac{1}{3}m_Z^2\cos 2\beta\sin^2\theta_W + m_b^2
  \end{array}\right)}\,.
\end{equation}
is diagonalized by the rotation matrix ${\cal R}^{\,\sb}$ such that
${\cal R}^{\,\sb\,\dagger}{\cal M}_{\sb}^2\, {\cal R}^{\,\sb} =
 {\rm diag}(\msb{1}^2,\,\msb{2}^2)$. ${\cal R}^{\,\sb}$ has the same structure like ${\cal R}^{\,\st}$
 given with (\ref{stopsmatrix}), and one can obtain it by making the interchange $\st \rightarrow \sb$.
\section{Interaction Lagrangian}

The part of the MSSM interaction Lagrangian used in our analytical calculations is given in this section.

The interaction of the charged Higgs boson with two quarks reads
\begin{equation}
{\cal L}_{H^\pm t b} =
  H^+\,\bar{t}\,(y_t\PL +y_b\PR )\,b + H^-\,\bar{b}\,(y_b\PL +y_t\PR )\,t\,,
\end{equation}
where the $\PL$ and $\PR$ are the left/ right projection
operators
\begin{equation}
 \PL = \frac{1}{2}(1-\g_5)\,, \quad
   \PR = \frac{1}{2}(1+\g_5)\,, \nonumber
\end{equation}
$y_t$ and $y_b$ are the tree-level couplings
\begin{equation}
  y_t = h_t\cos\b \,,\qquad
  y_b = h_b\sin\b \,,
\end{equation}
with $h_t$ and $h_b$ - the top and bottom Yukawa couplings
\begin{equation}
h_t=\frac{g}{\sqrt{2}m_W}\frac{1}{\sin\beta}\,, \qquad h_b=\frac{g}{\sqrt{2}m_W}\frac{1}{\cos\beta}\,.
\end{equation}

The interaction of two quarks with gluon exchange is given by
\begin{equation}
  \lag_{qqg} \,=\,
     -g_s\, T^a_{\! i j}\,G^a_{\!\mu}\,\bar{q}_i^{}\,\gamma^\mu\, q_j^{}\,,
\end{equation}
with  $i,j=1,2,3$ and $a=1,..,8$.

The interaction of the charged Higgs boson with two squarks is described by
\begin{equation}
{\cal L}_{H^\pm \st_i \sb_j} =
  (G_4^{})_{ij}^{}\,H^+\,\st_i^*\,\sb_j^{} +
  (G_4^*)_{ij}^{}\, H^-\,\sb_j^*\,\st_i^{} \,,
\end{equation}
with $i,j=1,2$,
\begin{equation}
G_4^{} = {\cal R}^{\,\st\,\dagger} \; \hat G_4^{} \; {\cal R}^{\,\sb}\,,
\label{eq:G4sq}
\end{equation}
and the matrix $\hat G_4$ is given by
\begin{equation}
\hspace*{-1cm}
\hat G_4^{ } = {\small
  \left(\! \begin{array}{cc}
    h_b m_b \sin\b + h_t m_t\cos\b - \rzw\,g\,m_W\sin\b\cos\b
    & h_b\,(A_b^*\sin\b + \mu\cos\b) \\[3mm]
      h_t\,(A_t\cos\b + \mu^*\sin\b) & h_t m_b \cos\b + h_b m_t \sin\b
\end{array}\! \right)} \,.
\label{eq:GLR4}
\end{equation}

The $W^\pm$-squark-squark interaction Lagrangian reads
\begin{eqnarray}
  \lag_{W^\pm \st_i \sb_j}=- \frac{ig}{\rzw}\,
  (\R_{Li}^{\st}\R_{Lj}^{\sb *}\,W^-_\mu\,\sb_j^*\delr\st_i^{} +
   \R_{Li}^{\st *}\R_{Lj}^{\sb}\,W^+_\mu\,\st_i^*\delr\sb_j^{})\,,
\label{eq:csqW}
\end{eqnarray}
where $i,j=1,2$ and $A\delr B = A\,(\partial_\mu B) - (\partial_\mu A)\,B.$

The squark-quark-gluino interaction is given by
\begin{eqnarray}
  {\cal L}_{q\sq\sg} =
  -\rzw\,g_s\,T_{jk}^a\left[\,
      \bar{\sg}_a ({\cal R}_{Li}^{\sq*}\,e^{-\frac{i}{2}\phi_3}\PL -
                  {\cal R}_{Ri}^{\sq*}\,e^{ \frac{i}{2}\phi_3}\PR)\,
      q^k\,\sq_{i}^{j *} \right.\nonumber \\
   \hspace{3cm}\left.
      +\,\bar q^{j} ({\cal R}_{Li}^{\sq}\,e^{ \frac{i}{2}\phi_3}\PR -
                   {\cal R}_{Ri}^{\sq}\,e^{-\frac{i}{2}\phi_3}\PL)\,\sg_{a}\,\sq_{i}^{k}
      \,\right]\,, \qquad
\end{eqnarray}
with $i=1,2$, $j,k=1,2,3$, and $a=1,..8$.

The interaction of two quarks with W-boson exchange is described by
\begin{equation}
  \lag_{W^\pm t b} \,=\, -\frac{g}{\rzw}\,
     (W^+_\mu\, \bar t\, \gamma^\mu\, \PL\, b +
      W^-_\mu\, \bar b\, \gamma^\mu\, \PL\, t)  \,.
\end{equation}
%
\section{Passarino-Veltman integrals}
The definitions of the Passarino--Veltman two-, and three-point functions~\cite{pave} in the convention
of~\cite{Denner} and the derived analytical expressions for their imaginary parts~\cite{HplusWh, Frank:Turan}
are given in this section.

The PV two-point functions are defined through the 4-dimensional integrals, as
\begin{eqnarray}
  B_0(p_1^2,m_0^2,m_1^2)   &=& \frac{1}{i\pi^2} \int d^{D}\! q \:
          \frac{1}{{\mathcal D}^0 {\mathcal D}^1} \,,\\
  B_\mu(p_1^2,m_0^2,m_1^2) &=& \frac{1}{i\pi^2} \int d^{D}\! q \:
      \frac{q_\mu}{{\mathcal D}^0 {\mathcal D}^1}
      = p_{1\mu}\, B_1(p_1^2,m_0^2,m_1^2) \,,
\end{eqnarray}
where we use the notation
\begin{equation}
  {\mathcal D}^{0} = q^{2} - m_{0}^{2}
  \quad \mbox{and}\quad
  {\mathcal D}^{j} = ( q + p_{j} )^{2} - m_{j}^{2}\,.
\end{equation}

In the rest frame system of
the (decaying) particle with impulse $p_1$, for the imaginary part
of $B_0$ we get
\begin{eqnarray} {\rm Im} \, B_0(M_1^2,m_0^2, m_1^2)=\frac{\pi
\lambda^{1/2}(M_1^2,m_0^2,m_1^2)}{M_1^2}\,, \label{first}
\end{eqnarray}
where the $\lambda$-function is defined as
\begin{eqnarray}
  \lambda (x,y,z)&=&
x^2+y^2+z^2 -2xy-2xz-2yz .\end{eqnarray}
In order to derive the imaginary part of $B_1$ we use the relation
\begin{eqnarray}
2 k^2 B_1(k^2, m_0^2, m_1^2)=A_0(m_0^2)-A_0(m_1^2)+(m_1^2-m_0^2-k^2)B_0(k^2, m_0^2, m_1^2)\,. \end{eqnarray}
Having in mind that ${\rm Im} \, A_0(m^2)=0$, we obtain
\begin{eqnarray} {\rm Im} \, B_1(M_1^2,m_0^2, m_1^2)=\frac{\pi (m_1^2-m_0^2-M_1^2)
\lambda^{1/2}(M_1^2,m_0^2,m_1^2)}{2 M_1^4}\,. \label{}
\end{eqnarray}
The PV three-point functions are defined as
\begin{eqnarray}
  C_0 (p_{1}^{2},(p_{1}-p_{2})^{2},p_{2}^{2},m_{0}^{2},m_{1}^{2},m_{2}^{2})&=& \frac{1}{i\pi^2} \int d^{D}\! q \:
          \frac{1}{{\mathcal D}^0 {\mathcal D}^1 {\mathcal D}^2} \,,  \nonumber \\
  C_\mu (p_{1}^{2},(p_{1}-p_{2})^{2},p_{2}^{2},m_{0}^{2},m_{1}^{2},m_{2}^{2})&=& \frac{1}{i\pi^2} \int d^{D}\! q \:
      \frac{q_\mu}{{\mathcal D}^0 {\mathcal D}^1 {\mathcal D}^2}
      = p_{1\mu} C_1 + p_{2\mu} C_2  \,. \nonumber \\ \label{PV3}
\end{eqnarray}
\begin{figure}[h!]
\begin{center}
 \resizebox{11cm}{!}{\includegraphics{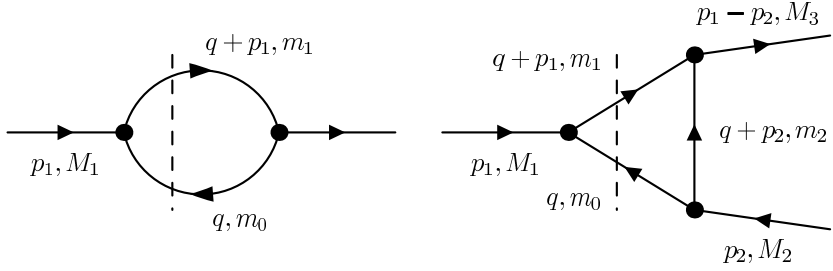}}
 \end{center}
\caption{The selfenergy and vertex type diagrams corresponding to the PV two- and three-point integrals,
when particles with masses $m_0$ and $m_1$ are on mass shell.}
 \label{fig:pv}
\end{figure}
For the absorptive parts of the integrals
 $C_0, C_1$ and $C_2$ in (\ref{PV3}) in the rest frame of the (decaying)
particle with momentum $p_1$, when particles with masses $m_0$ and $m_1$ are on mass shell, Fig.~\ref{fig:pv}, we obtain
\begin{eqnarray} {\rm Im} \,C_0(M_1^2, M_3^2, M_2^2, {\bf m_0^2, m_1^2},m_2^2)=
-\frac{\pi}{\lambda^{1/2}(M_1^2,M_2^2,M_3^2)}
\ln\vert\frac{a+b}{a-b}\vert \label{imc0}
\end{eqnarray}
 \begin{eqnarray} {\rm Im}\,C_1( M_1^2
,M_3^2,M_2^2,{\bf m_0^2, m_1^2},m_2^2)=\frac{M_2^2\, A -(p_1p_2)\,
B}{\Delta} \label{imc1} \end{eqnarray}
\begin{eqnarray}{\rm Im} \,C_2(M_1^2,
M_3^2, M_2^2, {\bf m_0^2, m_1^2},m_2^2)=-\frac{(p_1p_2)\,A -
M_1^2\, B}{\Delta}\label{imc2}
\end{eqnarray}
Here we have
\begin{eqnarray}
\Delta = M_1^2M_2^2 -(p_1p_2)^2\,,\qquad (p_1p_2)=\frac{M_1^2 +
M_2^2 -M_3^2}{2}\,, \qquad \nonumber \\
  a=
M_2^2+m_0^2 -m_2^2 + 2\,q^0p_2^0\,,\qquad b= -2 \,\vert\,
\vec{q}\,\vert\,\vert \vec{p_2}\,\vert\,, \qquad \qquad \nonumber \\
A=-\frac{\,\pi\,(m_1^2-m_0^2 -M_1^2)}{2\,
\lambda^{1/2}(M_1^2,M_2^2,M_3^2)}\, \ln
\vert\frac{a+b}{a-b}\vert\,, \qquad \qquad  \nonumber \\
B =\frac{\,\pi
}{\lambda^{1/2}(M_1^2,M_2^2,M_3^2)}\,
 \left\{ \frac{1}{2}(M_2^2+m_0^2 -m_2^2)
  \ln \vert\frac{a+b}{a-b}\vert
  + 2\, \vert\, \vec{q}\,\vert \,\vert \vec{p_2}\,\vert\right\},
\nonumber \\ q^0p_2^0=\frac{(m_1^2-m_0^2
-M_1^2)(M_1^2+M_2^2-M_3^2)}{4M_1^2}\,, \qquad \qquad \nonumber \\ \vert\,
\vec{q}\,\vert\,\vert
\vec{p_2}\,\vert=\frac{\lambda^{1/2}(M_1^2,m_0^2,m_1^2)\,\lambda^{1/2}(M_1^2,M_2^2,M_3^2)}{4M_1^2}\,.
\qquad \qquad \label{last}
\end{eqnarray}
In (\ref{imc0})-(\ref{imc2}) and further, we denote masses of the on-shell particles with bold font. Generally,
the functions $C_0, C_1$ and $C_3$ have three terms according to the three possible cuts over the particles in the
loops, and which of them is non zero depends on the kinematics.
The imaginary part of $C_0$ reads
\begin{eqnarray}
{\rm Im}\,C_0(M_1^2, M_3^2, M_2^2, m_0^2, m_1^2, m_2^2) = {\rm
Im}\,C_0(M_1^2, M_3^2, M_2^2,{\bf m_0^2, m_1^2}, m_2^2)+\nonumber \\
+{\rm Im}\,C_0(M_1^2, M_3^2, M_2^2, m_0^2,{\bf m_1^2, m_2^2})+{\rm Im}\,C_0(M_1^2, M_3^2, M_2^2, {\bf m_0^2}, m_1^2,
{\bf m_2^2})\,.
\end{eqnarray}
In terms of the analytic formulas (\ref{imc0})-(\ref{imc2}) we write
\begin{eqnarray}
{\rm Im}\,C_0(M_1^2, M_3^2, M_2^2, m_0^2, m_1^2, m_2^2)=
{\rm Im}\,C_0(M_1^2, M_3^2, M_2^2,{\bf m_0^2, m_1^2}, m_2^2)+\nonumber \\
+{\rm Im} \,C_0( M_3^2,M_1^2, M_2^2,m_2^2, {\bf  m_1^2,m_0^2})
+ {\rm Im} \,C_0(M_2^2, M_3^2, M_1^2, {\bf m_0^2},  m_2^2,{\bf
m_1^2}) \,.
\end{eqnarray}
\begin{eqnarray}
{\rm Im}\,C_1(M_1^2, M_3^2, M_2^2, m_0^2, m_1^2, m_2^2)={\rm
Im}\,C_1(M_1^2, M_3^2, M_2^2,{\bf m_0^2, m_1^2}, m_2^2)+\nonumber \\
+{\rm Im} \,C_1( M_3^2,M_1^2, M_2^2,m_2^2, {\bf  m_1^2,m_0^2})+
{\rm Im} \,C_2(M_2^2, M_3^2, M_1^2, {\bf m_0^2},  m_2^2,{\bf
m_1^2})\,,
\end{eqnarray}
\begin{eqnarray}
{\rm Im}\,C_2(M_1^2, M_3^2, M_2^2, m_0^2, m_1^2, m_2^2)={\rm
Im}\,C_2(M_1^2, M_3^2, M_2^2,{\bf m_0^2, m_1^2}, m_2^2)+\nonumber \\
+[{\rm Im} \,C_0( M_3^2,M_1^2, M_2^2,m_2^2, {\bf
m_1^2,m_0^2})+{\rm Im} \,C_1( M_3^2,M_1^2, M_2^2,m_2^2, {\bf
m_1^2,m_0^2})+\nonumber \\
+{\rm Im} \,C_2( M_3^2,M_1^2, M_2^2,m_2^2, {\bf m_1^2,m_0^2})]+
{\rm Im} \,C_1(M_2^2, M_3^2, M_1^2, {\bf m_0^2}, m_2^2,{\bf
m_1^2})\,.\, \end{eqnarray}

\end{appendix}



\begin{thebibliography}{99}

\bibitem{Dugan:1984qf} M.~Dugan, B.~Grinstein and L.~J.~Hall,
                       Nucl.\ Phys.\ B {\bf 255} (1985) 413.

\bibitem{Carena:1997ki} M.~Carena, M.~Quiros and C.~E.~Wagner,
                        Nucl.\ Phys.\ B {\bf 524} (1998) 3 [hep-ph/9710401];
                       for a review see: A.~G.~Cohen, D.~B.~Kaplan and A.~E.~Nelson,
                  Ann.\ Rev.\ Nucl.\ Part.\ Sci.\  {\bf 43} (1993) 27
                  [hep-ph/9302210].

\bibitem{Altarev:cf}
I.~S.~Altarev {\it et al.},
     Phys.\ Lett.\ B {\bf 276} (1992) 242;
I.~S.~Altarev {\it et al.},
     Phys.\ Atom.\ Nucl.\  {\bf 59} (1996) 1152
     [Yad.\ Fiz.\  {\bf 59N7} (1996) 1204];
E.~D.~Commins, S.~B.~Ross, D.~DeMille and B.~C.~Regan,
     Phys.\ Rev.\ A {\bf 50} (1994) 2960.

\bibitem{Nath:dn}
P.~Nath, Phys.\ Rev.\ Lett.\ {\bf 66} (1991) 2565;
Y.~Kizukuri and N.~Oshimo,
     Phys.\ Rev.\ D {\bf 46} (1992) 3025;
R.~Garisto and J.~D.~Wells,
     Phys.\ Rev.\ D {\bf 55} (1997) 1611 [hep-ph/9609511];
Y.~Grossman, Y.~Nir and R.~Rattazzi,
     Adv.\ Ser.\ Direct.\ High Energy Phys.\ {\bf 15} (1998) 755
     [hep-ph/9701231].

\bibitem{pilaftsis}
A. Pilaftsis, Phys. Rev. {\bf D 58} (1998)
096010 [hep-ph/9805373] and Phys.~Lett.~{\bf B 435} (1998) 88 [hep-ph/9805373;
A.~Pilaftsis and C.~E.~Wagner, Nucl.~Phys.~{\bf B 553}
(1999) 3 [hep-ph/9902371]; D. A. Demir, Phys.
Rev. {\bf D 60} (1999) 055006 [hep-ph/9901389].

\bibitem{carenaCP}
M. Carena, J. R. Ellis, A. Pilaftsis and C. E.
Wagner, Nucl. Phys. {\bf B 586} (2000) 92 [hep-ph/0003180].

\bibitem{Atwood:2000tu} For a review, see
                        D.~Atwood, S.~Bar-Shalom, G.~Eilam and A.~Soni,
                        Phys.\ Rept.\  {\bf 347} (2001) 1
                        [hep-ph/0006032].

\bibitem{Hplustb}
E. Christova, H. Eberl, E. Ginina, W. Majerotto, JHEP {\bf 0702}
(2007) 075 [hep-ph/0612088].

\bibitem{Hplustaunu}
E. Christova, H. Eberl, S. Kraml and W. Majerotto, JHEP {\bf 0212}
(2002) 021 [hep-ph/0211063].

\bibitem{HplusWh}
E.~Christova, E.~Ginina, M.~Stoilov JHEP {\bf 11}
(2003) 027 [hep-ph/0307319].



\bibitem{Hdecays}
E. Ginina, contribution to the 4th workshop "Gravity, Astrophysics, and Strings at the Black Sea", 2007, hep-ph/0801.2344

\bibitem{Katqproceedings}
E. Christova, H. Eberl, E. Ginina, arXiv:0812.0265 [hep-ph].

\bibitem{Jennifer} J. William, contributions to CPNSH Report, CERN-2006-009
(hep-ph/0608079)

\bibitem{jung:lee:song}
Kang Young Lee, Dong-Won Jung, H. S. Song,
Phys.\ Rev.\ D {\bf 70} (2004) 117701 [hep-ph/0307246].

\bibitem{Kidonakis}
N. Kidonakis, JHEP {\bf 0505}
(2005) 011, [hep-ph/0412422].

\bibitem{Hptb}
E. Christova, H. Eberl, S. Kraml and W. Majerotto,
     Nucl.\ Phys.\ B {\bf 639} (2002) 263;
E. Christova, H. Eberl, S. Kraml and W. Majerotto,
     Erratum to Nucl.\ Phys.\ B {\bf 639} (2002) 263.

\bibitem{FeynArts}
T. Hahn, Nucl. Phys. Proc. Suppl. B {\bf 89} (2000) 231; T. Hahn, {\it FeynArts User's Guide}, Comp.
Phys. Commun. {\bf 140} (2001) 418; T. Hahn, M. Perez-Victoria,
{\it FormCalc User's Guide}, Comput. Phys. Commun. {\bf 118} (1999) 153; T. Hahn, M. Perez-Victoria,
{\it LoopTools User's Guide}, Comput. Phys. Commun. {\bf 118} (1999) 153. (The software and all
manuals are available at http://www.feynarts.de.)
%
\bibitem{FFpackage}
G. J. van Oldenborgh, Comput. Phys. Commun. {\bf 66} (1991) 1.

%
\bibitem{PDFs}
J. Pumplin, D.R. Stump, J. Huston, H.L. Lai, Pavel M. Nadolsky,
W.K. Tung,  JHEP~{\bf 12} (2002) 0207 [hep-ph/0201195].
%

\bibitem{constraints}
M. Battaglia et al., Eur. Phys. J. C~{\bf 22} (2001) 535,
[arXiv:hep-ph/0106204].

\bibitem{stopsbotBR}
A. Bartl, K. Hidaka, Y. Kizukuri, T. Kon, W. Majerotto, Phys. Lett. B {\bf 315 } (1993) 360.

\bibitem{Belyaev:2002eq}
A.~Belyaev, D.~Garcia, J.~Guasch and J.~Sola, hep-ph/0203031.

\bibitem{pave} G.~Passarino and M.~J.~Veltman,
               Nucl.\ Phys.\ B {\bf 160} (1979) 151.


\bibitem{Denner}
A.~Denner, Fortschr. Phys. {\bf 41} (1993) 307.

\bibitem{Frank:Turan}
M. Frank, I. Turan, Phys.\ Rev.\ D {\bf 76} (2007) 016001, [hep-ph/0703184].

\end{thebibliography}
\end{document}